\documentclass[10pt,conference]{IEEEtran}
\usepackage{cite}
\usepackage{amsmath,amssymb,amsfonts}
\usepackage{algorithmic}
\usepackage{graphicx}
\usepackage{textcomp}
\usepackage{xcolor}
\usepackage[hyphens]{url}
\usepackage{fancyhdr}
\usepackage{hyperref}
\usepackage{tikz}
\usetikzlibrary{positioning,fit,backgrounds,calc,decorations.pathreplacing,patterns,intersections}
\usepackage{pgfplots}
\usepackage{pgfplotstable}
\usepgfplotslibrary{groupplots,statistics,fillbetween}
\pgfplotsset{compat=1.18}
\usepackage{booktabs}
\usepackage{listings}
\usepackage{subcaption}
\usepackage{enumitem}
\usepackage{tabularx}
\usepackage{booktabs}
\usepackage{makecell}

\newcommand{\cmark}{\checkmark}
\newcommand{\xmark}{$\times$}

\title{\LARGE{\scheme{}: What the Software Hides About LLMs Training Data, the Hardware Reveals at Scale, and Accelerators Magnify}}

\def\hpcacameraready{} 

\newcommand\hpcaauthors{Joshua Kalyanapu, Darsh Asher, Kaushal Mhapsekar, Bita Aslrousta, Rushiraj Chaitanyakumar Sheth,\\ Maharshi Mukeshkumar Oza, Achyuta Kannan, and Samira Mirbagher Ajorpaz}
\newcommand\hpcaaffiliation{North Carolina State University}
\newcommand\hpcaemail{}

\newcommand{\scheme}{\textsc{TranScope}}
\newcommand{\ctr}[1]{\textsl{#1}}
\newcommand{\raw}[1]{{\,\footnotesize\texttt{#1}}}

\renewcommand{\paragraph}[1]{\textbf{\emph{#1}}}
\usepackage{rotating}

\author{
  \ifdefined\hpcacameraready
    \IEEEauthorblockN{\hpcaauthors{}}
      \IEEEauthorblockA{
        \hpcaaffiliation{} \\
        \hpcaemail{}
      }
  \else
    \IEEEauthorblockN{\normalsize{HPCA \hpcayear{} Submission
      \textbf{\#\hpcasubmissionnumber{}}} \\
      \IEEEauthorblockA{
        Confidential Draft \\
        Do NOT Distribute!!
      }
    }
  \fi 
}

\fancypagestyle{camerareadyfirstpage}{%
  \fancyhead{}
  \renewcommand{\headrulewidth}{0pt}
  \fancyhead[C]{
    \ifdefined\aeopen
    \parbox[][12mm][t]{13.5cm}{\hpcayear{} IEEE International Symposium on High-Performance Computer Architecture (HPCA)}    
    \else
      \ifdefined\aereviewed
      \parbox[][12mm][t]{13.5cm}{\hpcayear{} IEEE International Symposium on High-Performance Computer Architecture (HPCA)}
      \else
      \ifdefined\aereproduced
      \parbox[][12mm][t]{13.5cm}{\hpcayear{} IEEE International Symposium on High-Performance Computer Architecture (HPCA)}
      \else
      \parbox[][0mm][t]{13.5cm}{\hpcayear{} IEEE International Symposium on High-Performance Computer Architecture (HPCA)}
    \fi 
    \fi 
    \fi 
    \ifdefined\aeopen 
      \includegraphics[width=12mm,height=12mm]{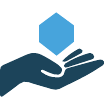}
    \fi 
    \ifdefined\aereviewed
      \includegraphics[width=12mm,height=12mm]{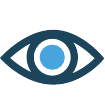}
    \fi 
    \ifdefined\aereproduced
      \includegraphics[width=12mm,height=12mm]{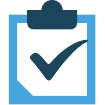}
    \fi
  }
  \fancyfoot[C]{}
}
\renewcommand{\headrulewidth}{0pt}

\begin{document}
\maketitle

\ifdefined\hpcacameraready 
  \thispagestyle{camerareadyfirstpage}
  \pagestyle{empty}
\else
  \thispagestyle{plain}
  \pagestyle{plain}
\fi

\newcommand{\hpcaheight}{0mm}
\ifdefined\eaopen
\renewcommand{\hpcaheight}{12mm}
\fi

\fancypagestyle{camerareadyfirstpage}{%
  \fancyhf{}%
  \renewcommand{\headrulewidth}{0pt}%
}
\fancypagestyle{camerareadypage}{%
  \fancyhf{}%
  \renewcommand{\headrulewidth}{0pt}%
}


\begin{abstract}

Membership is the root privacy primitive in machine learning: to date, no hardware-based out-of-distribution detection on black-box models has been demonstrated against constant-time, static neural networks with masked confidence. 

This paper performs the first cycle-level examination of how large language models (LLMs) and vision transformers interact with various modern microarchitecture components, including integrated accelerators, as LLMs scale in size and answers the question of whether the data that a model was trained on affects its execution footprint even without any input-dependent branch, dynamic optimization, or early exit and in constant-time models.  

The results confirm that the answer is yes and identify which modern hardware components, such as TLBs or on-core accelerators, reveal or amplify that effect. The results also answer whether the signal is informative enough to reliably classify the in-/vs/out-of-distribution property of membership. 


To understand why, we perform a systematic root cause analysis and find that the transformer's tokenization steps, which happen during training, alter the locality of the accesses the model makes to fetch the vocabulary token later during inference and, as a result, change the page table access patterns and TLB in a previously unknown data-dependent way, 
causing microarchitectural state to vary significantly based on whether or not the input was in the distribution of the transformer training data.

Building on the above observation, we introduce ~\scheme{}: the first microarchitecture tool for detecting membership information with low cost, no need for a surrogate model, and significantly higher robustness, e.g., 0.6 AUC for PETAL (best previously reported) vs 0.9 AUC (ours). This reintroduces hardware as both an opportunity, e.g., a tool for checking copyright violation for the first time, and a new channel for inferring membership (MIA). TranScope achieves a surprisingly high out-of-distribution detection accuracy of 98\% (1.5\% FPR, 2.5\% FNR) across large-scale models ranging from 4.4 million to 1.1 billion parameters.

\end{abstract}

\section{Introduction}


Membership inference attacks on generative LLMs have proven to be not as straightforward as in classifier models. Duan et al.~\cite{duan2024membership} shows how natural language lends itself to having a high degree of overlap between members and non-members, due to common phraseology appearing in natural language making members and non-members inherently similar. Meeus et al.~\cite{meeus2025sok} shows that most membership inference attacks on large-language models, particularly those benchmarking on the standard benchmark WikiMIA~\cite{shi2023detecting}, rely on distributional shifts between members and nonmembers, reducing recent LLM membership inference attacks to out-of-distribution detection instead. The equivalence is intuitive: from a safety standpoint, we are much more concerned with whether a user John Doe's social security number was used to train a model rather than whether the model was trained on the precise sentence "John Doe's social security number is X". Therefore for language models, we treat out-of-distribution detection as equivalent to membership inference attacks as in prior work~\cite{meeus2024did,shi2023detecting,he2025towards}.

\begin{table*}[t]
\centering
\small
\setlength{\tabcolsep}{6pt}
\setlength{\aboverulesep}{0pt}
\setlength{\belowrulesep}{0pt}
\setlength{\extrarowheight}{0.5ex}
\begin{tabular}{|l|c|c|c|c|c|c|c|c|c|c|c|c|c|c|c|c|c|c|}
\toprule
& \multicolumn{7}{c|}{\textbf{Requirements}}
& \multicolumn{5}{c|}{\textbf{Target}}
& \multicolumn{6}{c|}{\textbf{Mitigation}} \\
\midrule
\textbf{Work} &
\rotatebox[origin=l]{90}{Confidence} &
\rotatebox[origin=l]{90}{Output} &
\rotatebox[origin=l]{90}{Adaptive NN} &
\rotatebox[origin=l]{90}{Coresidency} &
\rotatebox[origin=l]{90}{White-box} &
\rotatebox[origin=l]{90}{Surrogate Model} &
\rotatebox[origin=l]{90}{One-shot input} &
\rotatebox[origin=l]{90}{LLM Membership} &
\rotatebox[origin=l]{90}{CNN Membership} &
\rotatebox[origin=l]{90}{MLP Membership} &
\rotatebox[origin=l]{90}{Model params} &
\rotatebox[origin=l]{90}{Victim data} &
\rotatebox[origin=l]{90}{Top-k logits} &
\rotatebox[origin=l]{90}{Rounded conf.} &
\rotatebox[origin=l]{90}{Label-only} &
\rotatebox[origin=l]{90}{Hard. update} &
\rotatebox[origin=l]{90}{Const. time} &
\rotatebox[origin=l]{90}{Diff. privacy} \\
\midrule
Shokri et al.~\cite{Shokri2017SP}
& \cmark & \cmark & \xmark & \xmark & $\circ$ & \cmark & \cmark & \xmark & \cmark & \cmark & \xmark & \xmark & \cmark & \cmark & \cmark & \cmark & \xmark & \cmark \\
TLBleed~\cite{gras_translation_2018}
& \xmark & \xmark & \xmark & \cmark & \xmark & \xmark & \xmark & \xmark & \xmark & \xmark & \xmark & \cmark & \xmark & \xmark & \xmark & \xmark & \xmark & \xmark \\
LiRA~\cite{carlini2022membership}
& \cmark & \cmark & \xmark & \xmark & \xmark & \cmark & \xmark & \cmark & \cmark & \cmark & \xmark & \xmark & \cmark & \cmark & \cmark & \xmark & \xmark & \cmark \\
Cache Telepathy~\cite{Yan2020CacheTelepathy}
& \xmark & \xmark & \xmark & \cmark & \xmark & \xmark & \xmark & \xmark & \cmark & \cmark & \cmark & \xmark & \xmark & \xmark & \xmark & \xmark & \xmark & \xmark \\
Akinsanya et al.~\cite{Akinsanya2024TimingCI}
& \xmark & \xmark & \cmark & \xmark & \xmark & \xmark & \xmark & \cmark & \cmark & \cmark & \xmark & \xmark & \xmark & \xmark & \xmark & \xmark & \cmark & \xmark \\
LOMIA~\cite{liu2026lomia}
& \xmark & \cmark & \xmark & \xmark & \xmark & \cmark & \xmark & \cmark & \xmark & \xmark & \xmark & \xmark & \xmark & \xmark & \xmark & \xmark & \xmark & \cmark \\
OSLO~\cite{peng2024oslo}
& \xmark & \cmark & \xmark & \xmark & \xmark & \cmark & \cmark & \xmark & \cmark & \xmark & \xmark & \xmark & \xmark & \xmark & \xmark & \xmark & \xmark & \xmark \\
GateBleed~\cite{kalyanapu2025gatebleed}
& \xmark & \xmark & \cmark & \cmark & \xmark & \xmark & \cmark & \cmark & \cmark & \cmark & \xmark & \cmark & \xmark & \xmark & \xmark & \cmark & \xmark & \xmark \\
Tong et al.~\cite{tong2025membership}
& \xmark & \xmark & \xmark & \xmark & \xmark & \cmark & \cmark & \cmark & \xmark & \xmark & \xmark & \xmark & \xmark & \xmark & \xmark & \xmark & \xmark & \xmark \\
PETAL~\cite{he2025towards}
& \xmark & \cmark & \xmark & \xmark & \xmark & \cmark & \xmark & \cmark & \xmark & \xmark & \xmark & \xmark & \xmark & \xmark & \xmark & \xmark & \xmark & \xmark \\
GhostTrace~\cite{srivastava2026ghosttrace}
& \xmark & \xmark & \xmark & \cmark & \xmark & \xmark & \cmark & \xmark & \cmark & \cmark & \xmark & \xmark & \xmark & \xmark & \xmark & \xmark & \xmark & \xmark \\
\midrule
\textbf{\scheme{} (this work)}
& \xmark & \xmark & \xmark & \cmark & \xmark & \xmark & \cmark & \cmark & \cmark & \xmark & \xmark & \xmark & \xmark & \xmark & \xmark & \xmark & \xmark & \xmark \\
\bottomrule
\end{tabular}
\caption{\scheme{} comparison to prior works.}
\label{tab:comparison-table}
\end{table*}

\begin{figure}[!t]
\centering
\begin{tikzpicture}

\pgfplotsset{
  defensepanel/.style={
    width=1.50cm, height=3.0cm, scale only axis,   
    ymin=45, ymax=105, ytick={50,60,70,80,90,100},
    xmin=0, xmax=3, xtick=\empty,
    title style={font=\scriptsize, yshift=-2pt},
    label style={font=\scriptsize}, tick label style={font=\tiny},
    ymajorgrids, grid style={line width=.2pt, draw=gray!25},
    axis line style={draw=gray!60}, tick style={draw=gray!60},
  },
  tradmark/.style={only marks, mark=o,       mark size=2.4pt, line width=.9pt},
  scmark/.style=  {only marks, mark=square*, mark size=2.4pt},
}
\newcommand{\chanceline}{%
  \addplot[dotted, line width=.7pt, domain=0:3, samples=2, forget plot] {50};}

\begin{groupplot}[
  defensepanel,
  group style={
    group size=4 by 1,
    horizontal sep=0.40cm,
    y descriptions at=edge left,   
  },
  ylabel={Attack accuracy (\%)},
]

\nextgroupplot[title={(a) None}]
  \chanceline
  \addplot[tradmark] coordinates {(1,92)};
  \addplot[scmark]   coordinates {(2,99)};

\nextgroupplot[title={(b) Top-$k$}]
  \chanceline
  \addplot[tradmark] coordinates {(1,83)};
  \addplot[scmark]   coordinates {(2,99)};

\nextgroupplot[title={(c) Rounded}]
  \chanceline
  \addplot[tradmark] coordinates {(1,66)};
  \addplot[scmark]   coordinates {(2,99)};

\nextgroupplot[title={(d) Label}]
  \chanceline
  \addplot[tradmark] coordinates {(1,53)};
  \addplot[scmark]   coordinates {(2,99)};
  \node[anchor=south east, font=\tiny, inner sep=1pt, yshift=0.5pt]
    at (axis cs:3,50) {chance};

\end{groupplot}

\node[anchor=north, yshift=-4pt, font=\scriptsize,
      draw=gray!60, line width=.4pt, rounded corners=2pt, inner sep=3pt]
  at (current bounding box.south) {%
    \tikz[baseline=-0.6ex]{\draw[line width=.9pt] (0,0) circle (2.4pt);}\;
    Traditional MIA\quad
    \tikz[baseline=-0.6ex]{\fill (-2.4pt,-2.4pt) rectangle (2.4pt,2.4pt);}\;
    \scheme{}};

\end{tikzpicture}
\caption{\scheme{} accuracy compared to the software MIA~\cite{Shokri2017SP} against machine learning, under confidence-hiding defenses: (a) none,
(b) top-$k$ logits, (c) rounded confidences, and (d) label-only outputs.
\scheme{} remains impervious to all software-level confidence-score-based
defenses~\cite{Shokri2017SP}; the dotted line marks chance accuracy.}
\label{fig:defense_comparison}
\end{figure}
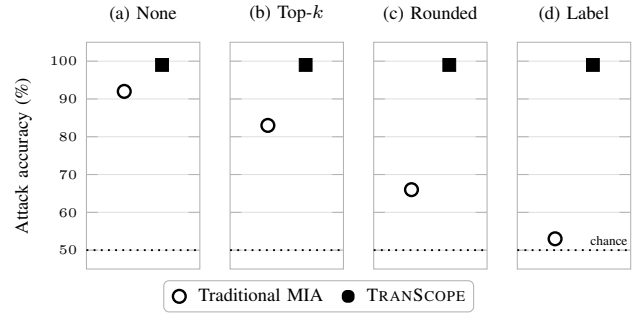
Traditional MIAs rely on confidence scores or logits.
The standard defense is output discretization---rounding, quantizing,
or binarizing confidence scores~\cite{Shokri2017SP}.
In production, logits are typically hidden, and many deployed MLaaS
systems expose only labels. \textbf{Figure~\ref{fig:defense_comparison}} shows how output discretization has a devastating effect on traditional software-level MIAs, while \scheme{} is unaffected by them since \scheme{} does not use model outputs or confidences. 

Recently, due to the proliferation of label-only model APIs research has shifted towards label-only membership inference attacks that do not rely on model internals or confidence scores. Instead, the model's robustness to an input, or how much noise is required to change the model output given a specific input, is measured via random perturbations and the robustness is used as a proxy for confidence, and therefore membership. Choquette-Choo et al.~\cite{choo2021membership} first presented a label-only attack on image classification models, finding comparable accuracy to confidence-based MIAs but required querying the expensive target model thousands of times.  OSLO~\cite{peng2024oslo} overcame this limitation with a one-shot membership inference attack, using trained surrogate models to precisely apply the minimum amount of noise such that a nonmember will flip classification, but the methodology does not extend to large language models and requires training surrogate models. PETAL~\cite{he2025towards} presented the first label-only membership inference attack on language models, but requires running a comparable-in-size surrogate model which is expensive. To date, no label-only membership inference attack has been demonstrated that do not require expensive running of a surrogate model, function on language models, and do not require multiple queries to the model.

\textcolor{black}{In concurrent work, Tong et al.~\cite{tong2025membership} showed how the training dataset is encoded in the tokenizer itself, using the presence of distinctive and domain-specific tokens in publicly available tokenizers to infer in-distribution data. Like \scheme{}, Tong et al. leak membership through the tokenizer rather than the model itself, but unlike \scheme{}, Tong et al. leaks through the presence of specific tokens whereas \scheme{} leaks through the inter-token distance inducing membership-dependent read strides on the hardware. Since \scheme{} does not depend on the presence of specific tokens in the tokenizer vocabulary but rather token distance strides, Tong et al.'s mitigation of adaptively removing distinctive tokens from the vocabulary is ineffective against \scheme{}.}

State-of-the-art membership inference attacks assume black-box access to AI models, where only a model's outputs are visible and the internals of the models are invisible which mirrors membership inference attacks against remote model APIs. However, AI companies are shifting towards binarizing models to run locally to adhere to data residency policies that restrict inputting sensitive information to a cloud AI model. The model weights and parameters constitute a large amount of the value of an AI company, prompting the AI company to release models in binarized form instead of open-weight form. Therefore, we anticipate in the future that AI model releases will be closed-weight, and model users will have only black-box access despite the model running on a machine that the user has root access to. Examples of deployed local inference models with only black-box access include Apple Foundation Models~\cite{gunter2024apple}, LLMWare's Model HQ~\cite{qualcommOnDeviceAgentic}, Nvidia TensorRT-compiled models~\cite{shafi2021demystifying}, Microsoft's Phi Silica~\cite{windowsSilicaSmall} and Aion Instruct~\cite{windowsExpandingOndevice}, and Adobe Firefly~\cite{banez2024extending}. Downloadable, closed-weight models that do not provide confidence scores disable confidence-based membership inference attacks and out-of-distribution detection force us to rely on expensive label-only membership inference attacks which are infeasible on low-power user hardware. 

The trend is clear - industry is moving towards binarized models, mitigating all white-box and confidence-score-based membership inference attacks and necessitating label-only attacks. Existing label-only attacks are expensive, requiring thousands of queries to a model, training a surrogate model, or running a surrogate model. \textbf{Table~\ref{tab:comparison-table}} positions \scheme{} within the literature, showing it as the first attack to not use confidence scores or outputs, not require an adaptive neural network, functions on language models, bypasses all existing MIA mitigations, does not require training or querying an expensive surrogate model, and only requires querying the target model once making it suitable for real-time deployment. 

\vspace{1em}
\noindent{\textbf{Limitations of prior hardware-assisted MIAs.}}
GateBleed (MICRO 2025)~\cite{kalyanapu2025gatebleed} was the first to exploit
hardware to bypass output-discretization defenses, achieving 81\%
accuracy via Intel AMX power-gating timing.
However, GateBleed rests on a structural assumption that disqualifies
it from the majority of modern systems: the target model must
contain an \emph{entropy-dependent branch} that causes different
computation for member versus nonmember inputs.
Adaptive networks such as BranchyNet~\cite{teerapittayanon2016branchynet}
and Mixture-of-Experts~\cite{zhou2022mixture} satisfy this condition,
but dominant modern models---LLaMA~\cite{touvron2023llama}, BERT~\cite{devlin2018bert}, GPT,
Vision Transformers---execute \emph{static}
forward passes, providing no input-dependent branch for GateBleed to
exploit.
Our measurements confirm that end-to-end inference time for GPT
and BERT is statistically indistinguishable between member and
nonmember samples to within 1.2\% across all model sizes and
microarchitectures---the timing or usage signal on which GateBleed and all prior hardware-based membership inference attacks depend simply does not exist in modern models.
Furthermore, GateBleed also requires an unpatched Intel AMX~\cite{intel2022amx},
present only on Sapphire Rapids through Granite Rapids CPUs, producing
no signal on earlier microarchitectures or non-Intel microarchitectures. GhostTrace~\cite{srivastava2026ghosttrace} presents a GPU performance-counter based membership inference attack on CNNs and ViT, but achieves lower accuracy than ~\scheme{} due to the relatively few performance counters exposed by GPUs as compared to CPUs.

Prior works EVAX~\cite{EVAX2022Micro}, PerSpectron~\cite{PerSpectron},
and RHMD~\cite{RHMD2017} demonstrated that malicious
computation from transient-execution attacks such as
Spectre~\cite{kocher2019spectre,canella2018systematic},
Meltdown~\cite{meltdown},
FallOut~\cite{canella2019fallout}, and
LVI~\cite{vanbulck2020lvi} leaves
microarchitectural footprints detectable through hardware
performance monitoring.
Our work flips this premise.
Instead of detecting microarchitectural side channels 
we use microarchitectural footprints to build a lightweight, fast detector for \emph{copyright abuse} detection and leak
properties of training data of production AI models including
GPT~\cite{radford2019language}, BERT~\cite{devlin2018bert}, and
Vision Transformers~\cite{dosovitskiy2020image}---
modern transformer architectures; not only with significantly higher robustness than software methods~\cite{tong2025membership} but also bypassing all confidence masking defenses~\cite{Shokri2017SP}. 

\definecolor{envline}{HTML}{0D366B}
\begin{figure}[t]
\centering
\begin{tikzpicture}
\begin{semilogxaxis}[
    width=7.05cm, height=3.80cm, scale only axis,
    xmin=3e6, xmax=4e9,
    ymin=70, ymax=103,
    xlabel={Parameter count},
    ylabel={Accuracy (\%)},
    log basis x=10,
    xtick={1e7,1e8,1e9},
    xticklabels={$10^7$,$10^8$,$10^9$},
    ytick={70,80,90,100},
    label style={font=\footnotesize}, tick label style={font=\scriptsize},
    grid=major, grid style={line width=.2pt, draw=gray!25},
    axis line style={draw=gray!60}, tick style={draw=gray!60},
]
\addplot[envline, solid, line width=1.1pt, mark=*, mark size=1.9pt,
         mark options={solid, fill=envline}]
coordinates {
    (4e6,84)     
    (11e6,89)    
    (29e6,100)   
    (41e6,100)   
    (70e6,99)    
    (110e6,99)   
    (160e6,99)   
    (340e6,100)  
    (1e9,94)     
    (1.4e9,97)   
    (2.8e9,99)   
};
\end{semilogxaxis}
\end{tikzpicture}
\caption{\scheme{} accuracy grows as LLMs scale. Each point is
the highest accuracy attains at that scale across all evaluated
model families.}
\label{fig:model_size_accuracy}
\end{figure}
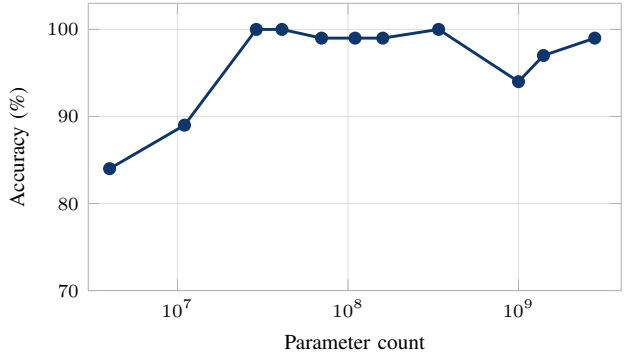

This paper demonstrates that the microarchitectural landscape can
leak \emph{training-data provenance}: whether a transformer or CNN
has been shaped by a specific sample or a copyright-relevant stylistic
property.
Privacy-critical training data reshapes the internal mechanics of
modern models---via embeddings, and
locality---into differences that are not merely statistically
detectable, but often \emph{visibly observable} in standard
microarchitectural counters (page-table walks, cache demand misses,
prefetch-hit behavior, pipeline clears).
Crucially, these leaks persist even when end-to-end timing changes by
less than 1.2\%, {bypassing constant-time defenses
entirely}---undermining the intuition that ``if the same
high-level inference code runs, the hardware footprint should be the
same.''
We establish the \emph{root cause} as a tokenizer-induced
spatial-locality gap: training changes which embedding rows are hot,
and that locality difference propagates into TLB walk counts, off-core
memory traffic, and reservation-station pressure while remaining
invisible to timing.

Even more critically, we show that microarchitectural training-data provenance leakage \emph{grows} with model size, reflecting the effect of larger model sizes memorizing more of the training set~\cite{carlini2022quantifying}. \textbf{Figure~\ref{fig:model_size_accuracy}} illustrates this point - a basic confidence-threshold membership inference attack jumps from 70\% accuracy to 96\% accuracy when the model surpasses 100M parameters. Our ~\scheme{} attacks on BERT and Pythia exhibit similar behavior. 

\begin{figure}[t]
\centering
\begin{tikzpicture}
\begin{axis}[
  ybar, bar width=6pt, width=0.96\columnwidth, height=4.6cm,
  enlarge x limits=0.18, ymin=0.965, ymax=1.080,
  ytick={0.97,0.99,1.01,1.03,1.05,1.07},
  yticklabel={$\pgfmathprintnumber[fixed,precision=2]{\tick}$},
  ylabel={Normalised power}, ylabel style={font=\small, yshift=2pt},
  xticklabel style={font=\small, align=center},
  xtick=data, symbolic x coords={tiny,mini,small,medium,base},
  xticklabels={4.4M,11M,29M,42M,110M},
  xticklabel style={font=\small, align=center},
  legend style={at={(0.02,0.99)},anchor=north west,font=\small,
    draw=black,thin,fill=white,cells={anchor=west},row sep=0pt},
  legend cell align=left, axis line style=thin, tick style=thin,
  clip=true,
]
\addplot[fill=black,draw=black] coordinates {
  (tiny,1.0)(mini,1.0028)(small,1.0009)(medium,1.0084)(base,1.0103)};
\addlegendentry{Member}
\addplot[fill=white,draw=black,
  postaction={pattern=north east lines,pattern color=black!60}] coordinates {
  (tiny,0.9935)(mini,1.0289)(small,1.0271)(medium,1.0485)(base,1.0560)};
\addlegendentry{Non-member}
\draw[densely dashed,gray!55,thin](axis cs:tiny,1.0)--(axis cs:base,1.0);
\node[font=\small,color=gray!80,anchor=south] at (axis cs:tiny,1.005){$-0.65\%^{\dagger}$};
\node[font=\small,color=black!60,anchor=south] at (axis cs:mini,1.026){$+2.6\%$};
\node[font=\small,color=black!60,anchor=south] at (axis cs:small,1.026){$+2.6\%$};
\node[font=\small,color=black!60,anchor=south] at (axis cs:medium,1.056){$+4.0\%$};
\node[font=\small,color=black!60,anchor=south] at (axis cs:base,1.065){$+4.5\%$};
\end{axis}
\end{tikzpicture}
\caption{Per-inference socket power normalised to BERT-tiny member.
Non-member exceeds member by 2.6--4.5\%, growing with model size.
${}^{\dagger}$Tiny: below RAPL quantum (artifact).}
\label{fig:power_normalised}
\end{figure}
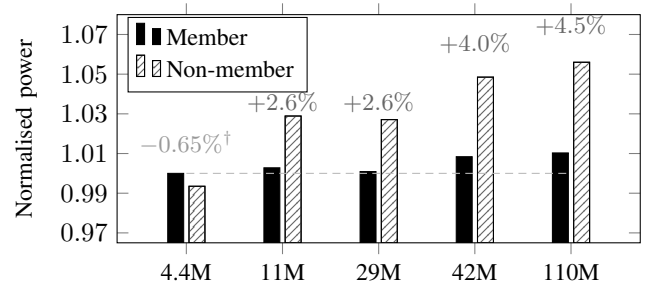

\textbf{Figure~\ref{fig:power_normalised}} additionally shows how power-consumption disparity between members and nonmembers, one statistic utilized by \scheme{}, grows with model size further highlighting the growing AI privacy threat with increasing model size. As AI models scale in size, \scheme{} is not going away but will only become stronger.

Unlike traditional microarchitectural side channels such as
Spectre~\cite{paulKocherSpectreAttacks}, Meltdown~\cite{meltdown}---which leak \emph{stored}
information (cryptographic keys, pointers, model
weights)---\scheme{} targets information \emph{not stored as bytes on
the machine}: training-set membership, a latent property of the
learned model previously not believed to be inferrable from hardware
alone.

\scheme{} is neither a TLBleed~\cite{gras_translation_2018} nor a traditional cache attack which encode victim memory in the TLB and cache for later extraction by an attacker --- while \scheme{} does use the TLB and cache, \scheme{} targets training-set membership, a property stored nowhere in victim memory. Similar to Cache Telepathy~\cite{Yan2020CacheTelepathy} which used known cache attacks to attack AI models, \scheme{} also attacks AI models; however, while Cache Telepathy targets model architecture and hyperparameters, \scheme{} targets training-set membership. This paper contains many unexpected results that are fundamental to hardware security in protecting user data used for training these models. 


\noindent{\textbf{Out-of-Distribution Detection and Membership Inference Attacks.}}
Out-of-distribution detection (OOD) asks whether a specific record is within the training-set distribution of a model; informally, whether the input is similar to the data the model was trained on. Out-of-distribution detection was first described by Hendrycks et al. in 2016~\cite{hendrycks2016baseline}, using model outputs due to models being overconfidently incorrect on inputs not drawn from its training distribution, leading to AI safety issues described by Amodei et al.~\cite{amodei2016concrete} which have manifested in the real world. In 2018, a pedestrian was killed by a self-driving Uber due to the software not recognizing the possibility of jaywalking pedestrians~\cite{NTSB2019TempeADS}. In 2016, a driver was killed when colliding with a tractor-trailer while using Tesla Autopilot, which failed to identify the tractor trailer crossing into the driver's lane and prompt for human intervention~\cite{NTSB2016Collision}. Thus, the inability of a model to identify inputs unlike its training set to flag a human for manual intervention can yield disastrous results. 

Related to out-of-distribution detection are membership inference attacks (MIAs), which ask whether a \emph{specific}, \emph{in-distribution} input was in the training set~\cite{Shokri2017SP,carlini2022membership}. Membership is a sensitive and legally actionable
fact---determining whether a cancer scan, private message, or
copyrighted manuscript was used for training can establish violations
of medical privacy, confidentiality agreements, or copyright.
Recent headlines make this concrete: Getty Images sued Stability AI
for training on copyrighted images~\cite{GettyImages_v_StabilityAI_2025_EWHC38_Ch}; LinkedIn was accused of training
on private messages; Apple was sued for allegedly using pirated books
to train AI models~\cite{Reuters2025AppleBooksAILawsuit}.
Foundational MIA work~\cite{Shokri2017SP,carlini2022membership} has
accumulated thousands of citations because membership is the root
privacy primitive: a model that behaves differently on training-set
members versus non-members enables escalation to attribute inference
and, in the worst case, model inversion.

\scheme{} removes both constraints simultaneously while achieving over 95\% membership inference accuracy in most cases.
By observing 290+ hardware performance counters rather than a single
timing scalar, it detects membership leakage in static,
branchless, production transformer networks---BERT~\cite{devlin2018bert}, ViT~\cite{wu2020visual},
GPT~\cite{radford2019language}, and Pythia~\cite{biderman2023pythia}---on Intel Sapphire Rapids, without requiring
any specialized accelerator or adaptive computation.
We demonstrate that 27+ counter events exhibit statistically
significant member-versus-nonmember divergence \emph{when timing is constant and every prior defense and attack is blind}.

      
      Our 290+ event traces reveal that LLMs and vision transformers
      do not stress the processor uniformly: training familiarity,
      input entropy, and model scale reshape branch behavior,
      front-end delivery, TLB activity, cache locality,
      reservation-station occupancy, execution-port pressure, power,
      and latency across the full pipeline.
      Each counter disparity is simultaneously a leakage signal and
      a design signal, motivating embedding-aware TLB partitioning,
      large-page-backed embedding tables, and branchless pooling
      primitives.
      
Legal cases involving copyrighted data in AI training have proliferated:
Thomson Reuters v.\ Ross~\cite{thomsonreuters2023},
Doe v.\ GitHub~\cite{doe1_2023},
Andersen v.\ Stability AI~\cite{andersen2023},
Getty Images v.\ Stability AI~\cite{GettyImages_v_StabilityAI_2025_EWHC38_Ch},
Concord Music v.\ Anthropic~\cite{concord2024},
UMG v.\ Suno~\cite{umg2025},
NYT v.\ OpenAI~\cite{grynbaum2023times},
Bartz v.\ Anthropic~\cite{bartz2025}, and
Kadrey v.\ Meta~\cite{kadrey2025}.
We validate \scheme{} on the same corpus at the center of the
\emph{New York Times Co.\ v.\ OpenAI.} litigation, using
real copyrighted NYT articles---not synthetic benchmarks---to
demonstrate that pre-2019 (potential member) and post-2019 (definite non-member) articles
separate in counter space despite identical software execution.

\vspace{1em}
\noindent \textbf{Contributions.} 
This paper makes the following contributions: 

\begin{itemize}[nosep]
\item We investigate the potential of  microarchitectural MIA on static,  branchless
      production transformer, bypassing constant-time defenses.
      We show that 15 Intel counter events and power consumption expose training-set
      membership in BERT, ViT, GPT, and LLaMA
yielding  AUC 0.995, 99.8\% accuracy, TPR\,100\%, FPR\,0.5\% on a realistic threat model available to common users of these popular models, 
      rendering confidence masking and constant-time padding ineffective.


\vspace{0.5em}
\item 
We trace the causal chain from training corpus vocabulary
       statistics through token-ID proximity and embedding-row spatial
      locality to cache/TLB residency and observable counter
       divergence and show the tokenizer effect on 
      embedding locality 
      affecting TLB is the root cause for LLM membership leakage.   

\vspace{0.5em}
\item  
      We show how ~\scheme{} bypasses software-level confidence masking and constant time padding for membership inference defenses due to \scheme{}'s lack of reliance on model outputs, confidences, and end-to-end execution time. As a first line-of-defense for mitigating ~\scheme{}, we recommend restricting tokenizer training during fine-tuning. Further defenses include smoothing embedding table accesses by sorting token IDs, making the spatial footprint of familiar and novel inputs more similar. We also introduce the concept of constant-footprint execution, where an AI inference system equalizes not only end-to-end runtime but other microarchitectural statistics like TLB accesses, execution port utilization, and per-inference energy consumption. 

\end{itemize}


\noindent\textbf{Responsible Disclosure.}
We disclosed \scheme{} to Intel in July 2026. Intel reproduced our results and confirmed our findings. 

\noindent\textbf{Paper organization.}
Section~\ref{sec:threat} enumerates the legal and malicious threat models under which we operate. Section~\ref{sec:attack} details the ~\scheme{} attack. Section~\ref{sec:eval} describes our experimental environment, including hardware, models, and datasets. Section~\ref{sec:arch} details the microarchitectural insights we glean to improve model execution from observing the performance counters generated by ~\scheme{}. 
Section~\ref{sec:related} discusses related work, and Section~\ref{sec:conclusion} concludes. 

\section{Threat Model}
\label{sec:threat}

We consider a legal threat model where ~\scheme{} imposes \emph{no requirements on the victim}: the attacker does not need to synchronise with, observe, or force the victim to repeat queries. In fact, there is no victim at all---the attacker, or legal auditor, \emph{is} the user.

\paragraph{Legal threat model (privileged auditor).}
The attacker downloads a publicly-available model and runs it on hardware
they own and control.
This is not a hypothetical scenario; it is the default operating mode
of every user who interacts with a publicly released language model
today.
Models such as GPT~\cite{radford2019language},
LLaMA~\cite{touvron2023llama}, and BERT~\cite{devlin2018bert} are
distributed as open weights executable on any commodity CPU without
specialized drivers, CUDA installation, or minimum VRAM. Furthermore, industry is shifting towards shipping binarized, closed-weight models runnable on commodity hardware, such as Apple Foundation Models~\cite{gunter2024apple}, Microsoft Aion Instruct~\cite{windowsExpandingOndevice}, and Adobe Firefly~\cite{banez2024extending}, allowing only black-box model accesses even when running on an attacker-controlled machine.
The attacker downloads the model, inputs the disputed
artifact to it, and reads the hardware counters on their own machine.
There is no co-location problem, no hypervisor to evade, no victim with which
to synchronize, and no need for the model owner to cooperate or
be aware that the measurement is taking place.
The attacker controls the entire measurement stack: the hardware, the
OS, the inference runtime, and the input.

This is the most realistic attacker model in the side-channel
literature, because it requires \emph{zero} capabilities beyond those
already available to every user who downloads a publicly-released model.
The goal is to build a hardware footprint that separates data matching
the model's training distribution from data that does not, producing
corroborating evidence for legal proceedings.

The CPU is the \emph{strongest} measurement platform for this purpose:
Intel Sapphire Rapids exposes 263 publicly documented core performance
monitoring events spanning branch prediction, front-end delivery,
TLB activity, cache hierarchy, scheduler pressure, execution-port
utilization, retirement, and software-prefetch behavior---orders of
magnitude more signal than GPU performance counters.






\paragraph{Auditor and expert-witness models.}
In forensic settings, a court-appointed or party-retained auditor
receives the artifact(s) under investigation, control corpora, and
controlled model access.
The deliverable is a reproducible evidence bundle satisfying
Federal Rule of Evidence~702 and the
\textit{Daubert} standard~\cite{hutchinson1993daubert}.
The \scheme{} protocol anticipates both plaintiff- and
defendant-assisted audits.

\paragraph{Empirical validation on real copyrighted articles.}
We apply \scheme{} to GPT on real New York Times article
snippets---copyrighted content from the NYT API, not synthetic
benchmarks.
We separate articles by GPT's disclosed training cutoff: pre-October~2019 (candidate members) versus post-2019 (definite non-members).
The results reveal statistically significant hardware footprint
differences that appear as a \emph{monotonic trend} over publication
date---precisely the signal a coarse aggregate histogram would
obscure.

\section{\scheme{}}
\label{sec:attack}

\subsection{Overview}

A model trained on a corpus $\mathcal{D}$ implicitly optimizes its
hardware access pattern for inputs drawn from $\mathcal{D}$.
Familiar inputs access a compact, hot region of the embedding table,
stressing the TLB and caches in predictable ways.
On the other hand, novel inputs scatter accesses across the embedding table, accessing a larger swath of the embedding table resulting in more TLB and cache misses. 
These effects are invisible at the level of the Python forward
pass---the same operators execute in the same order---yet are
measurable via hardware performance counters.

\scheme{} measures the \emph{distributional distance} between
an inference input and the model's training distribution, as reflected
in hardware telemetry.
This is why~\scheme{} is framed around in-distribution/out-of-distribution rather than
member/nonmember: it is a statistical characterization of whether the
input has the same ``spirit'' as the training corpus.

\paragraph{Provenance witness.}
We define a \emph{provenance witness} as a hardware-visible or
model-internal signal whose behavior is statistically consistent with
prior exposure to a disputed artifact, or to a narrowly defined
artifact class, and which can therefore serve as corroborating evidence
regarding the artifact's origin or training influence.
No single witness settles a case in isolation; when multiple witnesses
shift coherently---TLB activity, token locality, and power
all moving in the same direction---they collectively form a strong
provenance argument.

\subsection{Root Cause: Tokenizer-Induced Embedding Locality}
\label{subsec:embedding_locality}

\paragraph{Core claim.}
The counter divergence between familiar and unfamiliar inputs is not
noise across independent hardware events.
It is the projection of a \emph{single} microarchitectural mechanism
with a traceable causal chain:
\begin{center}
\emph{training-corpus familiarity $\;\rightarrow\;$ token-ID proximity
$\;\rightarrow\;$ embedding-row locality $\;\rightarrow\;$ cache/TLB
residency $\;\rightarrow\;$ counter divergence.}
\end{center}

\begin{figure}[!t]
    \centering
    \includegraphics[width=0.5\textwidth]{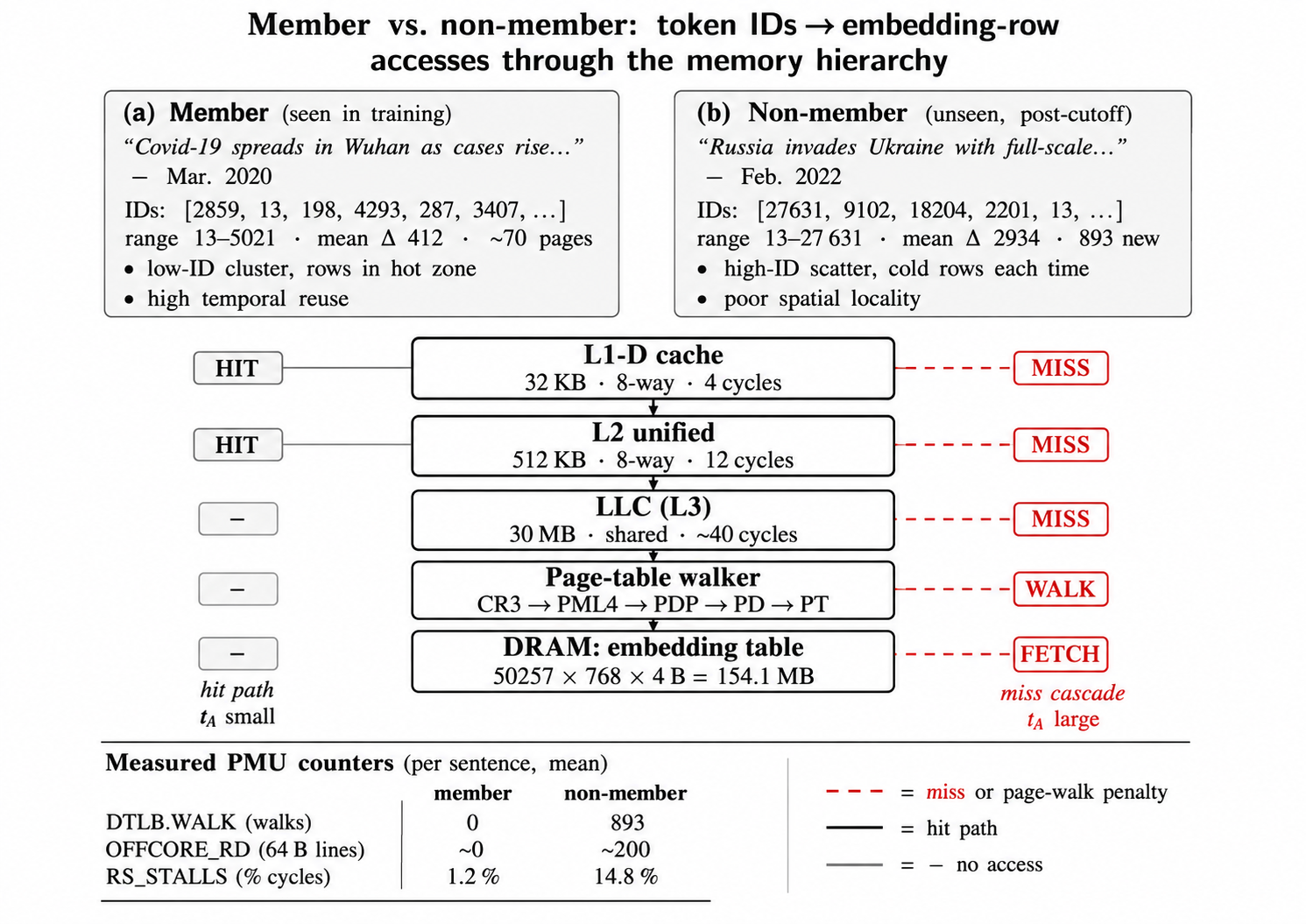}
    \caption{A \scheme{} root-cause mechanism: member vs.\ non-member
    inputs traverse identical software paths but diverge in the memory
    hierarchy.
    \textbf{Left:}\ A member article (COVID-19, Mar.\ 2020)
    produces low-ID, clustered tokens whose embedding rows are
    cache- and TLB-resident, yielding TLB hits and zero page-table walks.
    \textbf{Right:}\ A non-member article (Ukraine, Feb.\ 2022)
    produces high-ID, scattered tokens with large address strides,
    triggering the full TLB-miss $\to$ page-walk $\to$ DRAM-fetch
    cascade.
    The resulting counter divergence is consistent, large, and entirely
    invisible to end-to-end timing---bypassing constant-time defenses.}
    \label{fig:overview}
\end{figure}

We make this concrete with an example shown below and in \textbf{Figure~\ref{fig:overview}}. Suppose we have a large language model whose training cutoff was 2020. Then, if the model was trained on news articles around the cutoff time it would have encountered many headlines relating to COVID-19 (early 2020) but it would not have encountered any headlines relating to the Russo-Ukrainian war (early 2022). 

\paragraph{Two articles, two memory landscapes.}
Consider an example headline from March~2020: \emph{``COVID-19
spreads in Wuhan as cases rise; China reports new virus infections.''}
When the model's tokenizer processes this sentence, it maps each word to
an integer ID that serves as a row index into the model's embedding
table.
Because words like ``China,'' ``virus,'' and ``spread'' appeared
millions of times in the model's training corpus, the tokenizer learned to
assign them small, numerically nearby IDs: for example, \texttt{China\,(2859)},
\texttt{virus\,(3142)}, \texttt{spread\,(4293)}.
These IDs would cluster in a narrow band in the vocabulary. 

Now consider an example headline from February~2022: \emph{``Russia invades
Ukraine with full-scale attack, shelling near Kyiv.''}
The words ``Kyiv,'' ``invasion,'' and ``shelling'' would be rare or absent
in the model's training data.
The tokenizer would assign them large, widely spaced IDs, for example:
\texttt{Kyiv\,(27631)}, \texttt{invasion\,(18204)},
\texttt{shelling\,(14208)}.
These IDs scatter across a vast range of the vocabulary.

\definecolor{serA}{HTML}{86B6EF}
\definecolor{serB}{HTML}{5598E7}
\definecolor{serC}{HTML}{2A78D6}
\definecolor{serD}{HTML}{256ABF}
\definecolor{serE}{HTML}{184F95}
\definecolor{serF}{HTML}{0D366B}
\definecolor{zeroin}{HTML}{EB6834}

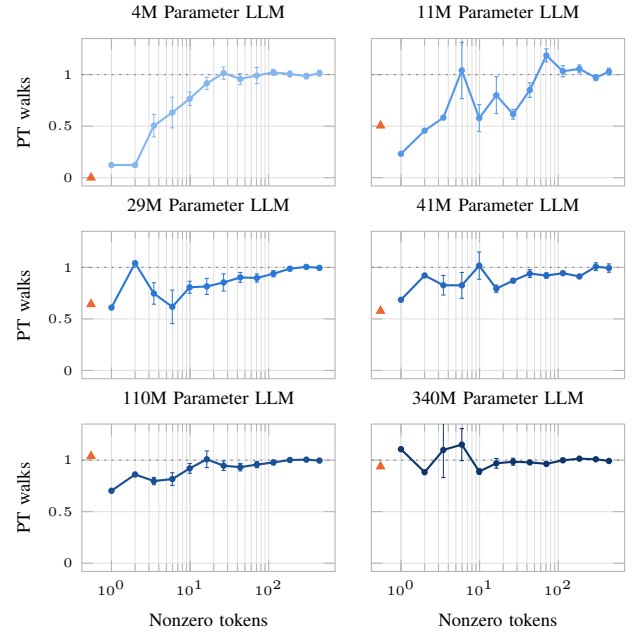
\begin{figure}[!t]
\centering
\begin{tikzpicture}
\pgfplotsset{
  sweep/.style={
    width=3.35cm, height=1.95cm, scale only axis,  
    xmode=log, xmin=0.42, xmax=700, ymin=-0.08, ymax=1.35,
    xtick={1,10,100}, ytick={0,0.5,1},
    label style={font=\scriptsize}, tick label style={font=\tiny},
    title style={font=\scriptsize, yshift=-2pt},
    grid=both, grid style={line width=.2pt, draw=gray!25},
    axis line style={draw=gray!60}, tick style={draw=gray!60},
    every axis plot/.append style={thick},
    clip mode=individual,
  },
  meas/.style={
    mark=*, mark size=.8pt,
    error bars/y dir=both, error bars/y explicit,
    error bars/error bar style={line width=.3pt},
  },
}
\newcommand{\sweepbase}{%
  \addplot[dotted,gray,line width=.4pt,forget plot]
    coordinates {(0.42,1) (700,1)};}
\newcommand{\sweepzero}[1]{%
  \addplot[only marks,mark=triangle*,mark size=1.4pt,color=zeroin]
    coordinates {(0.55,#1)};}

\begin{groupplot}[
  sweep,
  group style={
    group size=2 by 3,               
    horizontal sep=0.48cm, vertical sep=0.60cm,
    x descriptions at=edge bottom,   
    y descriptions at=edge left,     
  },
  xlabel={Nonzero tokens}, ylabel={PT walks},
]

\nextgroupplot[title={4M Parameter LLM}]
  \sweepbase
  \addplot[meas,color=serA] coordinates {
    (1,0.1227) +- (0,0.0000)
    (2,0.1227) +- (0,0.0000)
    (3.464,0.5046) +- (0,0.1091)
    (5.944,0.6319) +- (0,0.1486)
    (9.899,0.7665) +- (0,0.0653)
    (16.34,0.9155) +- (0,0.0590)
    (26.74,1.0145) +- (0,0.0597)
    (43.58,0.9579) +- (0,0.0525)
    (70.82,0.9908) +- (0,0.0780)
    (114.9,1.0224) +- (0,0.0287)
    (186.2,1.0063) +- (0,0.0282)
    (301.6,0.9856) +- (0,0.0264)
    (442.8,1.0144) +- (0,0.0286)
  };
  \sweepzero{0.0000}

\nextgroupplot[title={11M Parameter LLM}]
  \sweepbase
  \addplot[meas,color=serB] coordinates {
    (1,0.2317) +- (0,0.0000)
    (2,0.4565) +- (0,0.0000)
    (3.464,0.5829) +- (0,0.0176)
    (5.944,1.0405) +- (0,0.2732)
    (9.899,0.5779) +- (0,0.1313)
    (16.34,0.8006) +- (0,0.1790)
    (26.74,0.6158) +- (0,0.0476)
    (43.58,0.8502) +- (0,0.0703)
    (70.82,1.1864) +- (0,0.0627)
    (114.9,1.0323) +- (0,0.0535)
    (186.2,1.0559) +- (0,0.0383)
    (301.6,0.9710) +- (0,0.0287)
    (442.8,1.0290) +- (0,0.0331)
  };
  \sweepzero{0.5056}

\nextgroupplot[title={29M Parameter LLM}]
  \sweepbase
  \addplot[meas,color=serC] coordinates {
    (1,0.6100) +- (0,0.0000)
    (2,1.0410) +- (0,0.0000)
    (3.464,0.7468) +- (0,0.1042)
    (5.944,0.6172) +- (0,0.1630)
    (9.899,0.8069) +- (0,0.0583)
    (16.34,0.8152) +- (0,0.0781)
    (26.74,0.8536) +- (0,0.0833)
    (43.58,0.9026) +- (0,0.0480)
    (70.82,0.8969) +- (0,0.0400)
    (114.9,0.9383) +- (0,0.0304)
    (186.2,0.9871) +- (0,0.0208)
    (301.6,1.0046) +- (0,0.0173)
    (442.8,0.9954) +- (0,0.0192)
  };
  \sweepzero{0.6426}

\nextgroupplot[title={41M Parameter LLM}]
  \sweepbase
  \addplot[meas,color=serD] coordinates {
    (1,0.6853) +- (0,0.0000)
    (2,0.9216) +- (0,0.0000)
    (3.464,0.8266) +- (0,0.0950)
    (5.944,0.8256) +- (0,0.1253)
    (9.899,1.0169) +- (0,0.1323)
    (16.34,0.7936) +- (0,0.0359)
    (26.74,0.8710) +- (0,0.0212)
    (43.58,0.9397) +- (0,0.0381)
    (70.82,0.9209) +- (0,0.0259)
    (114.9,0.9434) +- (0,0.0178)
    (186.2,0.9118) +- (0,0.0149)
    (301.6,1.0070) +- (0,0.0376)
    (442.8,0.9930) +- (0,0.0405)
  };
  \sweepzero{0.5774}

\nextgroupplot[title={110M Parameter LLM}]
  \sweepbase
  \addplot[meas,color=serE] coordinates {
    (1,0.7027) +- (0,0.0000)
    (2,0.8608) +- (0,0.0000)
    (3.464,0.7970) +- (0,0.0352)
    (5.944,0.8163) +- (0,0.0616)
    (9.899,0.9198) +- (0,0.0469)
    (16.34,1.0082) +- (0,0.0811)
    (26.74,0.9458) +- (0,0.0466)
    (43.58,0.9320) +- (0,0.0363)
    (70.82,0.9566) +- (0,0.0301)
    (114.9,0.9774) +- (0,0.0219)
    (186.2,1.0014) +- (0,0.0194)
    (301.6,1.0046) +- (0,0.0153)
    (442.8,0.9954) +- (0,0.0160)
  };
  \sweepzero{1.0384}

\nextgroupplot[title={340M Parameter LLM}]
  \sweepbase
  \addplot[meas,color=serF] coordinates {
    (1,1.1061) +- (0,0.0000)
    (2,0.8827) +- (0,0.0000)
    (3.464,1.0986) +- (0,0.2674)
    (5.944,1.1505) +- (0,0.1556)
    (9.899,0.8897) +- (0,0.0275)
    (16.34,0.9690) +- (0,0.0472)
    (26.74,0.9843) +- (0,0.0336)
    (43.58,0.9780) +- (0,0.0163)
    (70.82,0.9637) +- (0,0.0189)
    (114.9,0.9984) +- (0,0.0190)
    (186.2,1.0141) +- (0,0.0137)
    (301.6,1.0084) +- (0,0.0125)
    (442.8,0.9916) +- (0,0.0127)
  };
  \sweepzero{0.9381}

\end{groupplot}
\end{tikzpicture}

\caption{2M/4M-sized page table walks for increasing LLM sizes with increasing zero-token count. Y-axis is page table walks normalized to page table walks for the fully random token sequence. 
}
\label{fig:bert-increasingly-member-figure}
\end{figure}

\begin{figure*}[!htbp]
  \centering
  \begin{subfigure}[b]{0.234\textwidth}
    \centering
\begin{tikzpicture}
\definecolor{idcurve}{HTML}{0D366B}
\definecolor{oodcurve}{HTML}{EB6834}
\begin{axis}[
    width=2.6cm, height=2.2cm, scale only axis,   
    xmin=0, xmax=30000, ymin=-0.02, ymax=1.04,
    xtick={0,15000,30000}, xticklabels={0,15k,30k},
    scaled x ticks=false,   
    ytick={0,0.5,1},
    ylabel={CDF}, ylabel style={yshift=-4pt},
    label style={font=\normalsize}, tick label style={font=\normalsize},
    grid=major, grid style={line width=.2pt, draw=gray!25},
    axis line style={draw=gray!60}, tick style={draw=gray!60},
]

\addplot[name path=id, idcurve, solid, line width=1.0pt, const plot mark right]
coordinates {
  (1,0.0079) (2,0.0079) (12,0.0394) (20,0.0394)
  (30,0.0866) (31,0.0866) (41,0.1181) (47,0.1181)
  (57,0.1496) (59,0.1496) (69,0.1732) (73,0.1732)
  (73,0.1811) (91,0.1811) (91,0.1890) (110,0.1890)
  (110,0.1968) (212,0.1968) (214,0.2126) (258,0.2126)
  (259,0.2283) (327,0.2283) (327,0.2362) (352,0.2362)
  (352,0.2441) (601,0.2441) (601,0.2520) (732,0.2520)
  (732,0.2598) (788,0.2598) (789,0.2756) (829,0.2756)
  (829,0.2835) (956,0.2835) (966,0.2992) (988,0.2992)
  (997,0.3543) (1011,0.3543) (1011,0.3701) (1023,0.3701)
  (1023,0.3858) (1056,0.3858) (1056,0.4016) (1067,0.4016)
  (1067,0.4094) (1099,0.4094) (1099,0.4173) (1172,0.4173)
  (1180,0.4409) (1186,0.4409) (1186,0.4488) (1323,0.4488)
  (1323,0.4646) (1376,0.4646) (1376,0.4724) (1474,0.4724)
  (1474,0.4803) (1535,0.4803) (1535,0.4882) (1745,0.4882)
  (1747,0.5039) (1845,0.5039) (1845,0.5118) (1933,0.5118)
  (1933,0.5197) (2148,0.5197) (2148,0.5354) (2234,0.5354)
  (2234,0.5433) (2416,0.5433) (2416,0.5512) (2760,0.5512)
  (2760,0.5591) (2807,0.5591) (2813,0.5748) (3260,0.5748)
  (3263,0.5906) (3348,0.5906) (3348,0.5984) (3417,0.5984)
  (3422,0.6142) (3620,0.6142) (3620,0.6220) (3651,0.6220)
  (3656,0.6378) (3674,0.6378) (3674,0.6457) (3893,0.6457)
  (3893,0.6535) (4047,0.6535) (4047,0.6614) (4239,0.6614)
  (4239,0.6693) (4653,0.6693) (4662,0.6850) (4681,0.6850)
  (4682,0.7008) (4701,0.7008) (4701,0.7087) (5111,0.7087)
  (5111,0.7165) (5534,0.7165) (5534,0.7244) (6935,0.7244)
  (6935,0.7323) (6996,0.7323) (6996,0.7402) (7270,0.7402)
  (7270,0.7480) (7484,0.7480) (7487,0.7638) (7530,0.7638)
  (7530,0.7716) (8495,0.7716) (8495,0.7795) (8850,0.7795)
  (8850,0.7874) (9002,0.7874) (9002,0.8032) (9511,0.8032)
  (9511,0.8110) (10273,0.8110) (10273,0.8189) (10341,0.8189)
  (10341,0.8268) (10368,0.8268) (10368,0.8347) (10385,0.8347)
  (10389,0.8504) (10431,0.8504) (10431,0.8583) (11128,0.8583)
  (11128,0.8661) (11140,0.8661) (11140,0.8740) (11376,0.8740)
  (11376,0.8819) (11399,0.8819) (11399,0.8898) (11851,0.8898)
  (11855,0.9134) (11925,0.9134) (11925,0.9213) (12922,0.9213)
  (12922,0.9291) (13729,0.9291) (13729,0.9370) (14923,0.9370)
  (14923,0.9449) (15946,0.9449) (15946,0.9528) (16236,0.9528)
  (16236,0.9606) (16556,0.9606) (16556,0.9685) (16936,0.9685)
  (16936,0.9764) (17408,0.9764) (17408,0.9842) (17506,0.9842)
  (17506,0.9921) (24383,0.9921) (24383,1.0000) (24383,1.0000)
  (30000,1.0000)
};

\addplot[name path=ood, oodcurve, densely dashed, line width=1.0pt,
         const plot mark right]
coordinates {
  (31,0.0079) (86,0.0079) (86,0.0158) (144,0.0158)
  (144,0.0236) (288,0.0236) (288,0.0315) (398,0.0315)
  (398,0.0394) (488,0.0394) (488,0.0472) (842,0.0472)
  (842,0.0551) (940,0.0551) (940,0.0630) (984,0.0630)
  (984,0.0709) (1050,0.0709) (1050,0.0787) (1229,0.0787)
  (1229,0.0866) (1484,0.0866) (1484,0.0945) (1625,0.0945)
  (1625,0.1102) (1725,0.1102) (1725,0.1181) (1746,0.1181)
  (1746,0.1260) (1787,0.1260) (1787,0.1339) (1872,0.1339)
  (1872,0.1417) (1992,0.1417) (1992,0.1496) (2011,0.1496)
  (2011,0.1575) (2069,0.1575) (2069,0.1653) (2237,0.1653)
  (2237,0.1732) (2443,0.1732) (2443,0.1811) (2672,0.1811)
  (2672,0.1890) (2719,0.1890) (2719,0.1968) (2751,0.1968)
  (2751,0.2047) (2862,0.2047) (2862,0.2126) (2983,0.2126)
  (2983,0.2205) (3026,0.2205) (3026,0.2283) (3066,0.2283)
  (3066,0.2362) (3344,0.2362) (3344,0.2441) (3391,0.2441)
  (3391,0.2520) (3408,0.2520) (3408,0.2598) (3447,0.2598)
  (3447,0.2677) (3618,0.2677) (3618,0.2756) (3711,0.2756)
  (3711,0.2835) (3802,0.2835) (3802,0.2913) (4108,0.2913)
  (4108,0.2992) (4223,0.2992) (4223,0.3071) (4395,0.3071)
  (4395,0.3150) (4449,0.3150) (4449,0.3228) (4797,0.3228)
  (4797,0.3307) (4956,0.3307) (4956,0.3386) (5401,0.3386)
  (5401,0.3465) (5533,0.3465) (5533,0.3543) (5815,0.3543)
  (5815,0.3622) (5924,0.3622) (5924,0.3701) (5941,0.3701)
  (5941,0.3780) (6033,0.3780) (6033,0.3858) (6062,0.3858)
  (6062,0.3937) (6077,0.3937) (6077,0.4016) (6974,0.4016)
  (6974,0.4094) (7306,0.4094) (7306,0.4173) (7450,0.4173)
  (7450,0.4252) (7640,0.4252) (7647,0.4409) (7734,0.4409)
  (7734,0.4488) (7801,0.4488) (7801,0.4567) (7974,0.4567)
  (7974,0.4646) (8233,0.4646) (8237,0.4803) (8314,0.4803)
  (8324,0.4961) (8446,0.4961) (8446,0.5039) (8481,0.5039)
  (8481,0.5118) (8910,0.5118) (8910,0.5197) (8964,0.5197)
  (8964,0.5276) (9023,0.5276) (9023,0.5354) (9295,0.5354)
  (9295,0.5433) (9741,0.5433) (9741,0.5512) (9963,0.5512)
  (9963,0.5591) (10238,0.5591) (10238,0.5669) (10376,0.5669)
  (10376,0.5748) (10436,0.5748) (10436,0.5827) (10693,0.5827)
  (10693,0.5906) (10826,0.5906) (10826,0.5984) (10855,0.5984)
  (10855,0.6063) (11228,0.6063) (11228,0.6142) (11628,0.6142)
  (11628,0.6220) (11922,0.6220) (11922,0.6299) (12280,0.6299)
  (12280,0.6378) (12604,0.6378) (12604,0.6457) (12746,0.6457)
  (12746,0.6535) (12768,0.6535) (12768,0.6614) (12830,0.6614)
  (12830,0.6693) (13268,0.6693) (13268,0.6772) (13347,0.6772)
  (13347,0.6850) (14026,0.6850) (14026,0.6929) (14232,0.6929)
  (14232,0.7008) (14282,0.7008) (14282,0.7087) (14435,0.7087)
  (14435,0.7165) (14538,0.7165) (14538,0.7244) (14590,0.7244)
  (14590,0.7323) (14753,0.7323) (14753,0.7402) (14764,0.7402)
  (14764,0.7480) (14776,0.7480) (14776,0.7559) (14801,0.7559)
  (14801,0.7638) (15455,0.7638) (15455,0.7716) (15474,0.7716)
  (15474,0.7795) (15505,0.7795) (15505,0.7874) (15558,0.7874)
  (15558,0.7953) (15884,0.7953) (15884,0.8032) (16127,0.8032)
  (16127,0.8110) (16157,0.8110) (16157,0.8189) (16308,0.8189)
  (16308,0.8268) (17205,0.8268) (17205,0.8347) (17778,0.8347)
  (17778,0.8425) (18100,0.8425) (18100,0.8504) (18980,0.8504)
  (18980,0.8583) (19238,0.8583) (19238,0.8661) (19483,0.8661)
  (19483,0.8740) (20624,0.8740) (20624,0.8819) (21004,0.8819)
  (21004,0.8898) (21065,0.8898) (21065,0.8976) (21092,0.8976)
  (21092,0.9055) (21466,0.9055) (21466,0.9134) (21498,0.9134)
  (21498,0.9213) (21801,0.9213) (21801,0.9291) (21986,0.9291)
  (21986,0.9370) (22146,0.9370) (22146,0.9449) (22397,0.9449)
  (22397,0.9528) (22418,0.9528) (22418,0.9606) (25229,0.9606)
  (25229,0.9685) (25836,0.9685) (25836,0.9764) (26652,0.9764)
  (26652,0.9842) (26753,0.9842) (26753,0.9921) (29258,0.9921)
  (29258,1.0000) (29258,1.0000) (30000,1.0000)
};


\end{axis}
\end{tikzpicture}
    \caption{4.4\,M}
  \end{subfigure}%
  \hfill%
  \begin{subfigure}[b]{0.19\textwidth}
    \centering
\begin{tikzpicture}
\definecolor{idcurve}{HTML}{0D366B}
\definecolor{oodcurve}{HTML}{EB6834}
\begin{axis}[
    width=2.6cm, height=2.2cm, scale only axis,   
    xmin=0, xmax=30000, ymin=-0.02, ymax=1.04,
    xtick={0,15000,30000}, xticklabels={0,15k,30k},
    scaled x ticks=false,   
    ytick={0,0.5,1},
    yticklabels={},   
    label style={font=\normalsize}, tick label style={font=\normalsize},
    grid=major, grid style={line width=.2pt, draw=gray!25},
    axis line style={draw=gray!60}, tick style={draw=gray!60},
]

\addplot[name path=id, idcurve, solid, line width=1.0pt, const plot mark right]
coordinates {
  (1,0.0079) (2,0.0079) (6,0.0315) (12,0.0315)
  (21,0.0551) (22,0.0551) (31,0.0945) (34,0.0945)
  (41,0.1181) (47,0.1181) (56,0.1417) (57,0.1417)
  (65,0.1653) (69,0.1653) (73,0.1811) (91,0.1811)
  (91,0.1890) (110,0.1890) (110,0.1968) (212,0.1968)
  (214,0.2126) (258,0.2126) (259,0.2283) (327,0.2283)
  (327,0.2362) (352,0.2362) (352,0.2441) (601,0.2441)
  (601,0.2520) (732,0.2520) (732,0.2598) (788,0.2598)
  (789,0.2756) (829,0.2756) (829,0.2835) (956,0.2835)
  (956,0.2913) (966,0.2913) (966,0.2992) (988,0.2992)
  (997,0.3543) (1011,0.3543) (1011,0.3701) (1023,0.3701)
  (1023,0.3858) (1056,0.3858) (1056,0.4016) (1067,0.4016)
  (1067,0.4094) (1099,0.4094) (1099,0.4173) (1172,0.4173)
  (1180,0.4409) (1186,0.4409) (1186,0.4488) (1323,0.4488)
  (1323,0.4646) (1376,0.4646) (1376,0.4724) (1474,0.4724)
  (1474,0.4803) (1535,0.4803) (1535,0.4882) (1745,0.4882)
  (1747,0.5039) (1845,0.5039) (1845,0.5118) (1933,0.5118)
  (1933,0.5197) (2148,0.5197) (2148,0.5354) (2234,0.5354)
  (2234,0.5433) (2416,0.5433) (2416,0.5512) (2760,0.5512)
  (2760,0.5591) (2807,0.5591) (2813,0.5748) (3260,0.5748)
  (3263,0.5906) (3348,0.5906) (3348,0.5984) (3417,0.5984)
  (3422,0.6142) (3620,0.6142) (3620,0.6220) (3651,0.6220)
  (3656,0.6378) (3674,0.6378) (3674,0.6457) (3893,0.6457)
  (3893,0.6535) (4047,0.6535) (4047,0.6614) (4239,0.6614)
  (4239,0.6693) (4653,0.6693) (4662,0.6850) (4681,0.6850)
  (4682,0.7008) (4701,0.7008) (4701,0.7087) (5111,0.7087)
  (5111,0.7165) (5534,0.7165) (5534,0.7244) (6935,0.7244)
  (6935,0.7323) (6996,0.7323) (6996,0.7402) (7270,0.7402)
  (7270,0.7480) (7484,0.7480) (7487,0.7638) (7530,0.7638)
  (7530,0.7716) (8495,0.7716) (8495,0.7795) (8850,0.7795)
  (8850,0.7874) (9002,0.7874) (9002,0.8032) (9511,0.8032)
  (9511,0.8110) (10273,0.8110) (10273,0.8189) (10341,0.8189)
  (10341,0.8268) (10368,0.8268) (10368,0.8347) (10385,0.8347)
  (10389,0.8504) (10431,0.8504) (10431,0.8583) (11128,0.8583)
  (11128,0.8661) (11140,0.8661) (11140,0.8740) (11376,0.8740)
  (11376,0.8819) (11399,0.8819) (11399,0.8898) (11851,0.8898)
  (11855,0.9134) (11925,0.9134) (11925,0.9213) (12922,0.9213)
  (12922,0.9291) (13729,0.9291) (13729,0.9370) (14923,0.9370)
  (14923,0.9449) (15946,0.9449) (15946,0.9528) (16236,0.9528)
  (16236,0.9606) (16556,0.9606) (16556,0.9685) (16936,0.9685)
  (16936,0.9764) (17408,0.9764) (17408,0.9842) (17506,0.9842)
  (17506,0.9921) (24383,0.9921) (24383,1.0000) (24383,1.0000)
  (30000,1.0000)
};

\addplot[name path=ood, oodcurve, densely dashed, line width=1.0pt,
         const plot mark right]
coordinates {
  (1,0.0079) (4,0.0079) (4,0.0158) (19,0.0158)
  (25,0.0394) (31,0.0394) (39,0.0630) (42,0.0630)
  (42,0.0709) (67,0.0709) (67,0.0866) (179,0.0866)
  (187,0.1024) (192,0.1024) (192,0.1102) (225,0.1102)
  (225,0.1181) (243,0.1181) (243,0.1260) (433,0.1260)
  (433,0.1339) (575,0.1339) (575,0.1417) (858,0.1417)
  (858,0.1496) (912,0.1496) (916,0.1653) (944,0.1653)
  (944,0.1732) (966,0.1732) (966,0.1811) (976,0.1811)
  (976,0.1890) (997,0.1890) (1001,0.2047) (1007,0.2047)
  (1011,0.2283) (1018,0.2283) (1018,0.2362) (1055,0.2362)
  (1062,0.2520) (1099,0.2520) (1099,0.2598) (1122,0.2598)
  (1122,0.2677) (1286,0.2677) (1294,0.2835) (1486,0.2835)
  (1486,0.2913) (1757,0.2913) (1757,0.2992) (1941,0.2992)
  (1941,0.3071) (1958,0.3071) (1964,0.3228) (2019,0.3228)
  (2019,0.3307) (2116,0.3307) (2116,0.3386) (2156,0.3386)
  (2156,0.3465) (2195,0.3465) (2195,0.3543) (2228,0.3543)
  (2228,0.3622) (2296,0.3622) (2296,0.3701) (2795,0.3701)
  (2795,0.3780) (2958,0.3780) (2958,0.3858) (3159,0.3858)
  (3159,0.3937) (3450,0.3937) (3450,0.4016) (3623,0.4016)
  (3623,0.4094) (3695,0.4094) (3695,0.4173) (4025,0.4173)
  (4025,0.4252) (4242,0.4252) (4242,0.4331) (4255,0.4331)
  (4255,0.4409) (4427,0.4409) (4427,0.4488) (4456,0.4488)
  (4456,0.4567) (4492,0.4567) (4492,0.4646) (4534,0.4646)
  (4534,0.4724) (4606,0.4724) (4606,0.4803) (5359,0.4803)
  (5359,0.4882) (5900,0.4882) (5900,0.4961) (5920,0.4961)
  (5920,0.5039) (6354,0.5039) (6354,0.5118) (6371,0.5118)
  (6371,0.5197) (6382,0.5197) (6382,0.5276) (6547,0.5276)
  (6547,0.5354) (7666,0.5354) (7666,0.5433) (7718,0.5433)
  (7718,0.5512) (7976,0.5512) (7976,0.5591) (8006,0.5591)
  (8006,0.5669) (8027,0.5669) (8027,0.5748) (8125,0.5748)
  (8125,0.5827) (8827,0.5827) (8827,0.5906) (9245,0.5906)
  (9245,0.5984) (9425,0.5984) (9425,0.6063) (9625,0.6063)
  (9625,0.6142) (10152,0.6142) (10152,0.6220) (10207,0.6220)
  (10207,0.6299) (10780,0.6299) (10780,0.6378) (11018,0.6378)
  (11018,0.6457) (11253,0.6457) (11253,0.6535) (11281,0.6535)
  (11281,0.6614) (11701,0.6614) (11701,0.6693) (11730,0.6693)
  (11730,0.6772) (11757,0.6772) (11757,0.6850) (12172,0.6850)
  (12172,0.6929) (13116,0.6929) (13116,0.7008) (13507,0.7008)
  (13507,0.7087) (13853,0.7087) (13853,0.7165) (14112,0.7165)
  (14112,0.7244) (14207,0.7244) (14207,0.7323) (14610,0.7323)
  (14610,0.7402) (14644,0.7402) (14644,0.7480) (15195,0.7480)
  (15195,0.7559) (15306,0.7559) (15306,0.7638) (15435,0.7638)
  (15435,0.7716) (15677,0.7716) (15677,0.7795) (15869,0.7795)
  (15869,0.7874) (15954,0.7874) (15954,0.7953) (16903,0.7953)
  (16905,0.8110) (17205,0.8110) (17205,0.8189) (17644,0.8189)
  (17644,0.8268) (18385,0.8268) (18385,0.8347) (18608,0.8347)
  (18608,0.8425) (18691,0.8425) (18691,0.8504) (18739,0.8504)
  (18739,0.8583) (19380,0.8583) (19380,0.8661) (19720,0.8661)
  (19720,0.8740) (20444,0.8740) (20444,0.8819) (21498,0.8819)
  (21498,0.8898) (22231,0.8898) (22231,0.8976) (22408,0.8976)
  (22408,0.9055) (23279,0.9055) (23279,0.9134) (23558,0.9134)
  (23558,0.9213) (23856,0.9213) (23863,0.9370) (24276,0.9370)
  (24276,0.9449) (24289,0.9449) (24289,0.9528) (24721,0.9528)
  (24721,0.9606) (25422,0.9606) (25422,0.9685) (25790,0.9685)
  (25790,0.9842) (28004,0.9842) (28004,0.9921) (28675,0.9921)
  (28675,1.0000) (28675,1.0000) (30000,1.0000)
};


\end{axis}
\end{tikzpicture}
    \caption{11.3\,M}
  \end{subfigure}%
  \hfill%
  \begin{subfigure}[b]{0.19\textwidth}
    \centering
\begin{tikzpicture}
\definecolor{idcurve}{HTML}{0D366B}
\definecolor{oodcurve}{HTML}{EB6834}
\begin{axis}[
    width=2.6cm, height=2.2cm, scale only axis,   
    xmin=0, xmax=30000, ymin=-0.02, ymax=1.04,
    xtick={0,15000,30000}, xticklabels={0,15k,30k},
    scaled x ticks=false,   
    ytick={0,0.5,1},
    yticklabels={},   
    label style={font=\normalsize}, tick label style={font=\normalsize},
    grid=major, grid style={line width=.2pt, draw=gray!25},
    axis line style={draw=gray!60}, tick style={draw=gray!60},
]

\addplot[name path=id, idcurve, solid, line width=1.0pt, const plot mark right]
coordinates {
  (1,0.0079) (2,0.0079) (6,0.0315) (12,0.0315)
  (20,0.0472) (21,0.0472) (28,0.0787) (30,0.0787)
  (35,0.1102) (41,0.1102) (49,0.1339) (56,0.1339)
  (59,0.1575) (65,0.1575) (73,0.1811) (91,0.1811)
  (91,0.1890) (110,0.1890) (110,0.1968) (212,0.1968)
  (214,0.2126) (258,0.2126) (259,0.2283) (327,0.2283)
  (327,0.2362) (352,0.2362) (352,0.2441) (601,0.2441)
  (601,0.2520) (732,0.2520) (732,0.2598) (788,0.2598)
  (789,0.2756) (829,0.2756) (829,0.2835) (956,0.2835)
  (956,0.2913) (966,0.2913) (966,0.2992) (988,0.2992)
  (993,0.3228) (997,0.3228) (997,0.3543) (1011,0.3543)
  (1011,0.3701) (1023,0.3701) (1023,0.3858) (1056,0.3858)
  (1056,0.4016) (1067,0.4016) (1067,0.4094) (1099,0.4094)
  (1099,0.4173) (1172,0.4173) (1180,0.4409) (1186,0.4409)
  (1186,0.4488) (1323,0.4488) (1323,0.4646) (1376,0.4646)
  (1376,0.4724) (1474,0.4724) (1474,0.4803) (1535,0.4803)
  (1535,0.4882) (1745,0.4882) (1747,0.5039) (1845,0.5039)
  (1845,0.5118) (1933,0.5118) (1933,0.5197) (2148,0.5197)
  (2148,0.5354) (2234,0.5354) (2234,0.5433) (2416,0.5433)
  (2416,0.5512) (2760,0.5512) (2760,0.5591) (2807,0.5591)
  (2813,0.5748) (3260,0.5748) (3263,0.5906) (3348,0.5906)
  (3348,0.5984) (3417,0.5984) (3422,0.6142) (3620,0.6142)
  (3620,0.6220) (3651,0.6220) (3656,0.6378) (3674,0.6378)
  (3674,0.6457) (3893,0.6457) (3893,0.6535) (4047,0.6535)
  (4047,0.6614) (4239,0.6614) (4239,0.6693) (4653,0.6693)
  (4653,0.6772) (4662,0.6772) (4662,0.6850) (4681,0.6850)
  (4682,0.7008) (4701,0.7008) (4701,0.7087) (5111,0.7087)
  (5111,0.7165) (5534,0.7165) (5534,0.7244) (6935,0.7244)
  (6935,0.7323) (6996,0.7323) (6996,0.7402) (7270,0.7402)
  (7270,0.7480) (7484,0.7480) (7487,0.7638) (7530,0.7638)
  (7530,0.7716) (8495,0.7716) (8495,0.7795) (8850,0.7795)
  (8850,0.7874) (9002,0.7874) (9002,0.8032) (9511,0.8032)
  (9511,0.8110) (10273,0.8110) (10273,0.8189) (10341,0.8189)
  (10341,0.8268) (10368,0.8268) (10368,0.8347) (10385,0.8347)
  (10389,0.8504) (10431,0.8504) (10431,0.8583) (11128,0.8583)
  (11128,0.8661) (11140,0.8661) (11140,0.8740) (11376,0.8740)
  (11376,0.8819) (11399,0.8819) (11399,0.8898) (11851,0.8898)
  (11855,0.9134) (11925,0.9134) (11925,0.9213) (12922,0.9213)
  (12922,0.9291) (13729,0.9291) (13729,0.9370) (14923,0.9370)
  (14923,0.9449) (15946,0.9449) (15946,0.9528) (16236,0.9528)
  (16236,0.9606) (16556,0.9606) (16556,0.9685) (16936,0.9685)
  (16936,0.9764) (17408,0.9764) (17408,0.9842) (17506,0.9842)
  (17506,0.9921) (24383,0.9921) (24383,1.0000) (24383,1.0000)
  (30000,1.0000)
};

\addplot[name path=ood, oodcurve, densely dashed, line width=1.0pt,
         const plot mark right]
coordinates {
  (10,0.0079) (11,0.0079) (16,0.0551) (22,0.0551)
  (27,0.0787) (39,0.0787) (47,0.1102) (106,0.1102)
  (108,0.1260) (130,0.1260) (130,0.1339) (156,0.1339)
  (161,0.1575) (196,0.1575) (196,0.1653) (219,0.1653)
  (219,0.1732) (396,0.1732) (396,0.1811) (552,0.1811)
  (552,0.1890) (621,0.1890) (621,0.1968) (858,0.1968)
  (864,0.2126) (911,0.2126) (911,0.2205) (920,0.2205)
  (920,0.2283) (935,0.2283) (943,0.2441) (964,0.2441)
  (964,0.2520) (979,0.2520) (986,0.2756) (993,0.2756)
  (1000,0.2992) (1007,0.2992) (1007,0.3071) (1024,0.3071)
  (1024,0.3150) (1033,0.3150) (1039,0.3386) (1042,0.3386)
  (1042,0.3465) (1054,0.3465) (1060,0.3622) (1164,0.3622)
  (1167,0.3780) (1367,0.3780) (1367,0.3858) (1396,0.3858)
  (1396,0.3937) (1448,0.3937) (1448,0.4016) (1529,0.4016)
  (1529,0.4094) (1562,0.4094) (1562,0.4173) (1669,0.4173)
  (1669,0.4252) (1915,0.4252) (1915,0.4331) (1956,0.4331)
  (1957,0.4488) (2118,0.4488) (2118,0.4567) (2738,0.4567)
  (2738,0.4646) (2844,0.4646) (2844,0.4724) (3008,0.4724)
  (3008,0.4803) (3041,0.4803) (3041,0.4882) (3202,0.4882)
  (3202,0.4961) (3399,0.4961) (3399,0.5039) (3962,0.5039)
  (3962,0.5118) (4409,0.5118) (4409,0.5197) (4602,0.5197)
  (4602,0.5276) (4713,0.5276) (4713,0.5354) (4833,0.5354)
  (4833,0.5433) (4872,0.5433) (4872,0.5512) (4978,0.5512)
  (4978,0.5591) (5166,0.5591) (5166,0.5669) (5238,0.5669)
  (5238,0.5748) (6262,0.5748) (6262,0.5827) (6333,0.5827)
  (6333,0.5906) (6792,0.5906) (6792,0.5984) (6816,0.5984)
  (6816,0.6063) (7254,0.6063) (7254,0.6142) (7447,0.6142)
  (7448,0.6299) (8394,0.6299) (8394,0.6378) (8794,0.6378)
  (8794,0.6457) (8881,0.6457) (8881,0.6535) (9004,0.6535)
  (9004,0.6614) (9015,0.6614) (9015,0.6693) (9221,0.6693)
  (9221,0.6772) (9543,0.6772) (9543,0.6850) (9582,0.6850)
  (9582,0.6929) (9654,0.6929) (9654,0.7008) (9793,0.7008)
  (9793,0.7087) (9867,0.7087) (9867,0.7165) (9919,0.7165)
  (9919,0.7244) (10444,0.7244) (10444,0.7323) (10782,0.7323)
  (10784,0.7480) (11195,0.7480) (11195,0.7559) (11753,0.7559)
  (11753,0.7638) (11962,0.7638) (11962,0.7716) (12662,0.7716)
  (12662,0.7795) (13353,0.7795) (13353,0.7874) (13506,0.7874)
  (13506,0.7953) (13635,0.7953) (13635,0.8032) (14794,0.8032)
  (14794,0.8110) (15028,0.8110) (15028,0.8189) (15354,0.8189)
  (15354,0.8268) (15607,0.8268) (15607,0.8347) (15853,0.8347)
  (15853,0.8425) (16336,0.8425) (16336,0.8504) (16665,0.8504)
  (16665,0.8583) (16934,0.8583) (16934,0.8661) (17594,0.8661)
  (17594,0.8740) (17965,0.8740) (17965,0.8819) (18035,0.8819)
  (18035,0.8898) (18119,0.8898) (18119,0.8976) (18288,0.8976)
  (18288,0.9055) (18319,0.9055) (18319,0.9134) (18860,0.9134)
  (18860,0.9213) (18875,0.9213) (18875,0.9291) (19008,0.9291)
  (19008,0.9370) (19401,0.9370) (19401,0.9449) (20463,0.9449)
  (20463,0.9528) (22359,0.9528) (22359,0.9606) (23499,0.9606)
  (23499,0.9685) (23644,0.9685) (23644,0.9764) (23860,0.9764)
  (23860,0.9842) (25941,0.9842) (25941,0.9921) (25982,0.9921)
  (25982,1.0000) (25982,1.0000) (30000,1.0000)
};


\end{axis}
\end{tikzpicture}
    \caption{29.1\,M}
  \end{subfigure}%
  \hfill%
  \begin{subfigure}[b]{0.19\textwidth}
    \centering
\begin{tikzpicture}
\definecolor{idcurve}{HTML}{0D366B}
\definecolor{oodcurve}{HTML}{EB6834}
\begin{axis}[
    width=2.6cm, height=2.2cm, scale only axis,   
    xmin=0, xmax=30000, ymin=-0.02, ymax=1.04,
    xtick={0,15000,30000}, xticklabels={0,15k,30k},
    scaled x ticks=false,   
    ytick={0,0.5,1},
    yticklabels={},   
    label style={font=\normalsize}, tick label style={font=\normalsize},
    grid=major, grid style={line width=.2pt, draw=gray!25},
    axis line style={draw=gray!60}, tick style={draw=gray!60},
]

\addplot[name path=id, idcurve, solid, line width=1.0pt, const plot mark right]
coordinates {
  (1,0.0079) (2,0.0079) (6,0.0315) (12,0.0315)
  (21,0.0551) (22,0.0551) (31,0.0945) (34,0.0945)
  (41,0.1181) (47,0.1181) (56,0.1417) (57,0.1417)
  (65,0.1653) (69,0.1653) (73,0.1811) (91,0.1811)
  (91,0.1890) (110,0.1890) (110,0.1968) (212,0.1968)
  (214,0.2126) (258,0.2126) (259,0.2283) (327,0.2283)
  (327,0.2362) (352,0.2362) (352,0.2441) (601,0.2441)
  (601,0.2520) (732,0.2520) (732,0.2598) (788,0.2598)
  (789,0.2756) (829,0.2756) (829,0.2835) (956,0.2835)
  (956,0.2913) (966,0.2913) (966,0.2992) (988,0.2992)
  (997,0.3543) (1011,0.3543) (1011,0.3701) (1023,0.3701)
  (1023,0.3858) (1056,0.3858) (1056,0.4016) (1067,0.4016)
  (1067,0.4094) (1099,0.4094) (1099,0.4173) (1172,0.4173)
  (1180,0.4409) (1186,0.4409) (1186,0.4488) (1323,0.4488)
  (1323,0.4646) (1376,0.4646) (1376,0.4724) (1474,0.4724)
  (1474,0.4803) (1535,0.4803) (1535,0.4882) (1745,0.4882)
  (1747,0.5039) (1845,0.5039) (1845,0.5118) (1933,0.5118)
  (1933,0.5197) (2148,0.5197) (2148,0.5354) (2234,0.5354)
  (2234,0.5433) (2416,0.5433) (2416,0.5512) (2760,0.5512)
  (2760,0.5591) (2807,0.5591) (2813,0.5748) (3260,0.5748)
  (3263,0.5906) (3348,0.5906) (3348,0.5984) (3417,0.5984)
  (3422,0.6142) (3620,0.6142) (3620,0.6220) (3651,0.6220)
  (3656,0.6378) (3674,0.6378) (3674,0.6457) (3893,0.6457)
  (3893,0.6535) (4047,0.6535) (4047,0.6614) (4239,0.6614)
  (4239,0.6693) (4653,0.6693) (4662,0.6850) (4681,0.6850)
  (4682,0.7008) (4701,0.7008) (4701,0.7087) (5111,0.7087)
  (5111,0.7165) (5534,0.7165) (5534,0.7244) (6935,0.7244)
  (6935,0.7323) (6996,0.7323) (6996,0.7402) (7270,0.7402)
  (7270,0.7480) (7484,0.7480) (7487,0.7638) (7530,0.7638)
  (7530,0.7716) (8495,0.7716) (8495,0.7795) (8850,0.7795)
  (8850,0.7874) (9002,0.7874) (9002,0.8032) (9511,0.8032)
  (9511,0.8110) (10273,0.8110) (10273,0.8189) (10341,0.8189)
  (10341,0.8268) (10368,0.8268) (10368,0.8347) (10385,0.8347)
  (10389,0.8504) (10431,0.8504) (10431,0.8583) (11128,0.8583)
  (11128,0.8661) (11140,0.8661) (11140,0.8740) (11376,0.8740)
  (11376,0.8819) (11399,0.8819) (11399,0.8898) (11851,0.8898)
  (11855,0.9134) (11925,0.9134) (11925,0.9213) (12922,0.9213)
  (12922,0.9291) (13729,0.9291) (13729,0.9370) (14923,0.9370)
  (14923,0.9449) (15946,0.9449) (15946,0.9528) (16236,0.9528)
  (16236,0.9606) (16556,0.9606) (16556,0.9685) (16936,0.9685)
  (16936,0.9764) (17408,0.9764) (17408,0.9842) (17506,0.9842)
  (17506,0.9921) (24383,0.9921) (24383,1.0000) (24383,1.0000)
  (30000,1.0000)
};

\addplot[name path=ood, oodcurve, densely dashed, line width=1.0pt,
         const plot mark right]
coordinates {
  (0,0.0079) (3,0.0394) (15,0.0394) (15,0.0551)
  (64,0.0551) (65,0.0709) (77,0.0709) (77,0.0787)
  (101,0.0787) (101,0.0866) (143,0.0866) (143,0.0945)
  (173,0.0945) (173,0.1024) (215,0.1024) (215,0.1102)
  (228,0.1102) (228,0.1181) (381,0.1181) (381,0.1260)
  (436,0.1260) (436,0.1339) (738,0.1339) (738,0.1417)
  (793,0.1417) (793,0.1496) (961,0.1496) (961,0.1575)
  (1014,0.1575) (1014,0.1653) (1029,0.1653) (1029,0.1732)
  (1187,0.1732) (1187,0.1811) (1306,0.1811) (1306,0.1890)
  (1461,0.1890) (1461,0.1968) (1476,0.1968) (1476,0.2047)
  (1886,0.2047) (1886,0.2126) (1988,0.2126) (1988,0.2205)
  (2200,0.2205) (2201,0.2362) (2237,0.2362) (2237,0.2441)
  (2248,0.2441) (2248,0.2520) (2270,0.2520) (2270,0.2598)
  (2284,0.2598) (2284,0.2677) (2543,0.2677) (2543,0.2756)
  (2630,0.2756) (2630,0.2835) (2743,0.2835) (2744,0.2992)
  (3570,0.2992) (3570,0.3071) (4129,0.3071) (4129,0.3150)
  (4205,0.3150) (4205,0.3228) (4252,0.3228) (4252,0.3307)
  (4288,0.3307) (4288,0.3386) (4677,0.3386) (4677,0.3465)
  (4802,0.3465) (4802,0.3543) (4862,0.3543) (4862,0.3622)
  (4874,0.3622) (4874,0.3701) (4939,0.3701) (4939,0.3780)
  (5062,0.3780) (5062,0.3858) (5126,0.3858) (5126,0.3937)
  (5763,0.3937) (5772,0.4094) (6484,0.4094) (6484,0.4173)
  (6614,0.4173) (6614,0.4252) (6858,0.4252) (6858,0.4331)
  (6961,0.4331) (6961,0.4646) (6974,0.4646) (6974,0.4961)
  (7018,0.4961) (7018,0.5039) (7094,0.5039) (7094,0.5118)
  (7619,0.5118) (7619,0.5197) (8041,0.5197) (8041,0.5276)
  (8275,0.5276) (8275,0.5354) (8757,0.5354) (8757,0.5433)
  (8875,0.5433) (8875,0.5512) (8951,0.5512) (8951,0.5591)
  (9127,0.5591) (9127,0.5669) (9320,0.5669) (9320,0.5748)
  (10742,0.5748) (10742,0.5827) (10837,0.5827) (10837,0.5906)
  (11222,0.5906) (11222,0.5984) (11282,0.5984) (11282,0.6063)
  (11573,0.6063) (11573,0.6142) (11599,0.6142) (11599,0.6220)
  (11805,0.6220) (11805,0.6299) (12511,0.6299) (12511,0.6378)
  (12733,0.6378) (12742,0.6535) (12760,0.6535) (12760,0.6614)
  (13866,0.6614) (13866,0.6693) (13911,0.6693) (13911,0.6772)
  (13935,0.6772) (13935,0.6850) (14109,0.6850) (14109,0.6929)
  (14155,0.6929) (14155,0.7008) (14363,0.7008) (14363,0.7087)
  (14775,0.7087) (14775,0.7165) (14809,0.7165) (14811,0.7323)
  (15170,0.7323) (15170,0.7402) (15383,0.7402) (15383,0.7480)
  (15495,0.7480) (15495,0.7559) (17647,0.7559) (17647,0.7638)
  (17660,0.7638) (17660,0.7716) (18109,0.7716) (18109,0.7795)
  (18541,0.7795) (18541,0.7874) (18869,0.7874) (18874,0.8032)
  (18883,0.8032) (18883,0.8110) (19315,0.8110) (19315,0.8189)
  (20010,0.8189) (20010,0.8268) (20129,0.8268) (20129,0.8347)
  (20173,0.8347) (20173,0.8425) (20228,0.8425) (20228,0.8504)
  (21298,0.8504) (21306,0.8661) (22070,0.8661) (22070,0.8740)
  (22198,0.8740) (22198,0.8819) (22442,0.8819) (22442,0.8898)
  (23157,0.8898) (23157,0.8976) (23468,0.8976) (23468,0.9055)
  (24535,0.9055) (24535,0.9134) (24668,0.9134) (24668,0.9213)
  (24711,0.9213) (24711,0.9291) (25070,0.9291) (25070,0.9370)
  (25341,0.9370) (25341,0.9449) (25508,0.9449) (25508,0.9528)
  (25981,0.9528) (25981,0.9606) (26987,0.9606) (26987,0.9685)
  (27403,0.9685) (27403,0.9764) (27753,0.9764) (27753,0.9842)
  (28685,0.9842) (28685,0.9921) (28809,0.9921) (28809,1.0000)
  (28809,1.0000) (30000,1.0000)
};


\end{axis}
\end{tikzpicture}
    \caption{41.7\,M}
  \end{subfigure}%
  \hfill%
  \begin{subfigure}[b]{0.19\textwidth}
    \centering
\begin{tikzpicture}
\definecolor{idcurve}{HTML}{0D366B}
\definecolor{oodcurve}{HTML}{EB6834}
\begin{axis}[
    width=2.6cm, height=2.2cm, scale only axis,   
    xmin=0, xmax=30000, ymin=-0.02, ymax=1.04,
    xtick={0,15000,30000}, xticklabels={0,15k,30k},
    scaled x ticks=false,   
    ytick={0,0.5,1},
    yticklabels={},   
    label style={font=\normalsize}, tick label style={font=\normalsize},
    grid=major, grid style={line width=.2pt, draw=gray!25},
    axis line style={draw=gray!60}, tick style={draw=gray!60},
]

\addplot[name path=id, idcurve, solid, line width=1.0pt, const plot mark right]
coordinates {
  (1,0.0079) (2,0.0079) (6,0.0315) (12,0.0315)
  (21,0.0551) (22,0.0551) (31,0.0945) (34,0.0945)
  (41,0.1181) (47,0.1181) (56,0.1417) (57,0.1417)
  (65,0.1653) (69,0.1653) (73,0.1811) (91,0.1811)
  (91,0.1890) (110,0.1890) (110,0.1968) (212,0.1968)
  (214,0.2126) (258,0.2126) (259,0.2283) (327,0.2283)
  (327,0.2362) (352,0.2362) (352,0.2441) (601,0.2441)
  (601,0.2520) (732,0.2520) (732,0.2598) (788,0.2598)
  (789,0.2756) (829,0.2756) (829,0.2835) (956,0.2835)
  (956,0.2913) (966,0.2913) (966,0.2992) (988,0.2992)
  (997,0.3543) (1011,0.3543) (1011,0.3701) (1023,0.3701)
  (1023,0.3858) (1056,0.3858) (1056,0.4016) (1067,0.4016)
  (1067,0.4094) (1099,0.4094) (1099,0.4173) (1172,0.4173)
  (1180,0.4409) (1186,0.4409) (1186,0.4488) (1323,0.4488)
  (1323,0.4646) (1376,0.4646) (1376,0.4724) (1474,0.4724)
  (1474,0.4803) (1535,0.4803) (1535,0.4882) (1745,0.4882)
  (1747,0.5039) (1845,0.5039) (1845,0.5118) (1933,0.5118)
  (1933,0.5197) (2148,0.5197) (2148,0.5354) (2234,0.5354)
  (2234,0.5433) (2416,0.5433) (2416,0.5512) (2760,0.5512)
  (2760,0.5591) (2807,0.5591) (2813,0.5748) (3260,0.5748)
  (3263,0.5906) (3348,0.5906) (3348,0.5984) (3417,0.5984)
  (3422,0.6142) (3620,0.6142) (3620,0.6220) (3651,0.6220)
  (3656,0.6378) (3674,0.6378) (3674,0.6457) (3893,0.6457)
  (3893,0.6535) (4047,0.6535) (4047,0.6614) (4239,0.6614)
  (4239,0.6693) (4653,0.6693) (4662,0.6850) (4681,0.6850)
  (4682,0.7008) (4701,0.7008) (4701,0.7087) (5111,0.7087)
  (5111,0.7165) (5534,0.7165) (5534,0.7244) (6935,0.7244)
  (6935,0.7323) (6996,0.7323) (6996,0.7402) (7270,0.7402)
  (7270,0.7480) (7484,0.7480) (7487,0.7638) (7530,0.7638)
  (7530,0.7716) (8495,0.7716) (8495,0.7795) (8850,0.7795)
  (8850,0.7874) (9002,0.7874) (9002,0.8032) (9511,0.8032)
  (9511,0.8110) (10273,0.8110) (10273,0.8189) (10341,0.8189)
  (10341,0.8268) (10368,0.8268) (10368,0.8347) (10385,0.8347)
  (10389,0.8504) (10431,0.8504) (10431,0.8583) (11128,0.8583)
  (11128,0.8661) (11140,0.8661) (11140,0.8740) (11376,0.8740)
  (11376,0.8819) (11399,0.8819) (11399,0.8898) (11851,0.8898)
  (11855,0.9134) (11925,0.9134) (11925,0.9213) (12922,0.9213)
  (12922,0.9291) (13729,0.9291) (13729,0.9370) (14923,0.9370)
  (14923,0.9449) (15946,0.9449) (15946,0.9528) (16236,0.9528)
  (16236,0.9606) (16556,0.9606) (16556,0.9685) (16936,0.9685)
  (16936,0.9764) (17408,0.9764) (17408,0.9842) (17506,0.9842)
  (17506,0.9921) (24383,0.9921) (24383,1.0000) (24383,1.0000)
  (30000,1.0000)
};

\addplot[name path=ood, oodcurve, densely dashed, line width=1.0pt,
         const plot mark right]
coordinates {
  (2,0.0079) (11,0.0472) (12,0.0472) (15,0.0787)
  (22,0.0787) (24,0.0945) (110,0.0945) (110,0.1024)
  (177,0.1024) (177,0.1102) (249,0.1102) (249,0.1181)
  (343,0.1181) (343,0.1260) (417,0.1260) (417,0.1339)
  (491,0.1339) (491,0.1417) (690,0.1417) (690,0.1496)
  (701,0.1496) (701,0.1575) (722,0.1575) (722,0.1653)
  (733,0.1653) (738,0.1811) (988,0.1811) (988,0.1968)
  (1014,0.1968) (1014,0.2047) (1027,0.2047) (1027,0.2126)
  (1043,0.2126) (1050,0.2362) (1063,0.2362) (1064,0.2520)
  (1423,0.2520) (1423,0.2598) (1574,0.2598) (1574,0.2677)
  (1709,0.2677) (1709,0.2756) (1720,0.2756) (1720,0.2835)
  (1995,0.2835) (1995,0.2913) (2215,0.2913) (2215,0.2992)
  (2344,0.2992) (2344,0.3071) (2408,0.3071) (2408,0.3150)
  (2476,0.3150) (2476,0.3228) (2694,0.3228) (2694,0.3307)
  (2769,0.3307) (2769,0.3386) (2938,0.3386) (2938,0.3465)
  (3027,0.3465) (3027,0.3543) (3174,0.3543) (3174,0.3622)
  (3195,0.3622) (3199,0.3780) (3532,0.3780) (3532,0.3858)
  (3757,0.3858) (3757,0.3937) (3824,0.3937) (3824,0.4016)
  (3931,0.4016) (3931,0.4094) (3968,0.4094) (3968,0.4173)
  (3979,0.4173) (3979,0.4252) (4095,0.4252) (4095,0.4331)
  (4156,0.4331) (4156,0.4409) (4202,0.4409) (4202,0.4488)
  (4234,0.4488) (4234,0.4567) (4487,0.4567) (4487,0.4646)
  (4660,0.4646) (4660,0.4724) (4846,0.4724) (4846,0.4803)
  (4860,0.4803) (4860,0.4882) (5931,0.4882) (5931,0.4961)
  (5953,0.4961) (5953,0.5039) (6020,0.5039) (6020,0.5118)
  (6379,0.5118) (6379,0.5197) (6776,0.5197) (6776,0.5276)
  (6961,0.5276) (6961,0.5433) (6974,0.5433) (6974,0.5591)
  (7094,0.5591) (7094,0.5669) (7254,0.5669) (7254,0.5827)
  (8078,0.5827) (8078,0.5906) (8346,0.5906) (8346,0.5984)
  (8852,0.5984) (8852,0.6063) (9199,0.6063) (9199,0.6142)
  (9464,0.6142) (9464,0.6220) (10019,0.6220) (10019,0.6299)
  (10244,0.6299) (10244,0.6378) (10709,0.6378) (10709,0.6457)
  (10748,0.6457) (10748,0.6535) (11408,0.6535) (11408,0.6614)
  (11574,0.6614) (11574,0.6693) (11612,0.6693) (11612,0.6772)
  (11812,0.6772) (11812,0.6850) (11957,0.6850) (11957,0.6929)
  (12043,0.6929) (12043,0.7008) (12054,0.7008) (12054,0.7087)
  (12417,0.7087) (12417,0.7165) (12429,0.7165) (12429,0.7244)
  (12502,0.7244) (12502,0.7323) (12924,0.7323) (12924,0.7402)
  (13179,0.7402) (13179,0.7480) (13338,0.7480) (13338,0.7559)
  (13403,0.7559) (13403,0.7638) (13798,0.7638) (13798,0.7716)
  (14031,0.7716) (14031,0.7795) (14714,0.7795) (14714,0.7874)
  (14727,0.7874) (14727,0.7953) (15204,0.7953) (15204,0.8032)
  (15666,0.8032) (15666,0.8110) (17274,0.8110) (17274,0.8189)
  (18273,0.8189) (18273,0.8268) (18351,0.8268) (18351,0.8347)
  (18683,0.8347) (18683,0.8425) (18856,0.8425) (18856,0.8504)
  (19475,0.8504) (19475,0.8583) (20344,0.8583) (20344,0.8661)
  (20450,0.8661) (20450,0.8740) (20535,0.8740) (20535,0.8819)
  (21093,0.8819) (21093,0.8898) (21149,0.8898) (21149,0.8976)
  (21193,0.8976) (21193,0.9055) (22103,0.9055) (22103,0.9134)
  (22153,0.9134) (22153,0.9213) (22504,0.9213) (22504,0.9291)
  (23915,0.9291) (23916,0.9449) (24535,0.9449) (24535,0.9528)
  (25472,0.9528) (25472,0.9606) (26102,0.9606) (26102,0.9685)
  (26673,0.9685) (26673,0.9764) (27236,0.9764) (27236,0.9842)
  (28014,0.9842) (28014,0.9921) (28279,0.9921) (28279,1.0000)
  (28279,1.0000) (30000,1.0000)
};


\end{axis}
\end{tikzpicture}
    \caption{110.1\,M}
  \end{subfigure}
  \caption{Successive token-distance CDF for member vs.\ nonmember across model sizes---root-cause observation that member tokens cluster while nonmember tokens scatter regardless of model size, and evidence that the phenomenon generalizes across sizes and is not limited to one case. The dark curves are an in-distribution sample and the dashed orange curves are an out-of-distribution sample.}
  \label{fig:token_distance_cdfs}
\end{figure*}
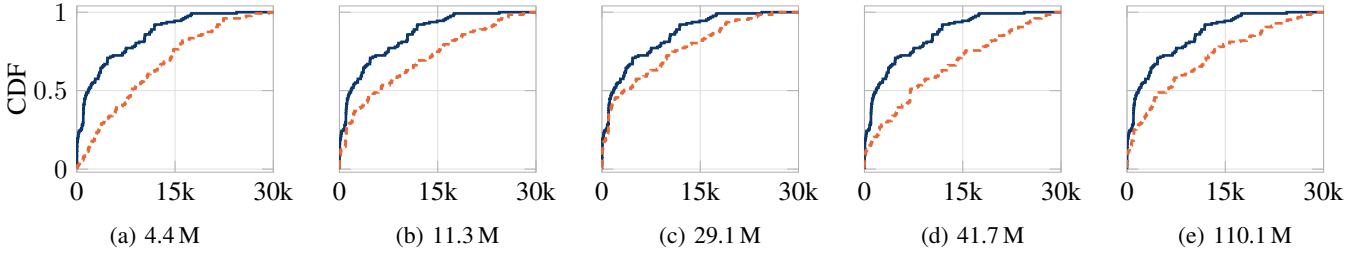

\paragraph{From IDs to memory addresses.}
Each token ID $r_t$ indexes a row in the embedding table at address
$\texttt{addr}(r_t) = \texttt{BASE} + r_t \cdot d_{\text{model}}
\cdot \texttt{sizeof(float)}$.
For BERT-base ($d_{\text{model}} = 768$, fp32), each row occupies
$768 \times 4 = 3{,}072$\,bytes---nearly filling a single 4\,KB
memory page.
The physical distance the processor must jump between successive
token lookups is proportional to the difference in their IDs.
For the member pair China\,$\to$\,virus, the stride is
$283 \times 3{,}072 \approx 0.83$\,MB---a short hop within
a few neighbouring memory pages.
For the non-member pair Kyiv\,$\to$\,invasion, the stride is
$9{,}427 \times 3{,}072 \approx 28.9$\,MB---a 35$\times$ larger
jump that leaps across thousands of pages the processor has never
seen in this inference.

\paragraph{What the hardware sees.}
The CPU maintains a translation lookaside buffer (TLB) that caches
recent virtual-to-physical page mappings.
When successive embedding lookups land on nearby pages, the TLB
already holds the translation: the access is a fast hit, resolved in a few cycles.
When lookups scatter across distant pages, each new page forces a
\emph{page-table walk}---the processor must traverse four levels of
the page-table hierarchy in memory, costing roughly 200 cycles per
walk.

\textbf{Figure~\ref{fig:bert-increasingly-member-figure}} makes this concrete. In this experiment, we created a dataset consisting of token sequences of length 512, beginning as entirely random sequences and setting an increasing amount of tokens within the sequence to 0 until the last item in the dataset, an all-zero token sequence. We run the dataset through all canonical BERT sizes, finding that tiny, mini, small, medium, and base exhibit a clear trend of increasing page table walks for increasing amount of nonzero tokens. Large stays flat at a constant number of page table walks, we attribute this to the large size of the model drowning out the contribution of the embedding table accesses. However, in Section~\ref{sec:eval} we find that many performance counter observations can be combined in aggregate to determine out-of-distribution from in-distribution data at high accuracy, even for large models.

For the candidate member article, the clustered token IDs produce a compact
memory footprint.
The embedding rows land on $\leq\!70$ distinct pages, and most of
those pages stay TLB-resident throughout inference.
The processor follows the fast path: TLB hit, L1/L2 cache hit, no
page-table walk.

For the non-member article, the scattered IDs spray accesses across
many different pages.
The TLB overflows, and each new page triggers the slow path: TLB miss,
second-level TLB miss, last-level cache miss, page-table walk, and
finally a DRAM fetch.

The model executes an \emph{identical} sequence of neural-network
operations in the \emph{same} number of clock cycles for both articles---the difference in page-table walks is too small (20 walks out of 52 million
cycles) to surface in end-to-end timing, 
yet the hardware counters that track TLB misses, off-core memory
traffic, and reservation-station pressure tell different
stories.
This is the signal \scheme{} reads and why constant-time
padding provides no defense.


\definecolor{oodc}{HTML}{EB6834}
\definecolor{idc}{HTML}{2A78D6}

\pgfplotsset{
  bxaxis/.style={
    scale only axis,
    xmin=0.4, xmax=2.6,
    xtick={1,2}, xticklabels={OOD,ID},
    boxplot/draw direction=y,
    boxplot/box extend=0.62,
    boxplot/whisker extend=0.34,
    boxplot/every box/.style={line width=0.5pt},
    boxplot/every whisker/.style={line width=0.5pt, solid},
    boxplot/every median/.style={line width=1.1pt},
    label style={font=\scriptsize}, tick label style={font=\tiny},
    title style={font=\scriptsize, yshift=-2pt},
    ymajorgrids, grid style={line width=.2pt, draw=gray!25},
    axis line style={draw=gray!60}, tick style={draw=gray!60},
    xtick style={draw=none},
    clip=false,
  },
  bxood/.style={draw=oodc, fill=oodc!12},
  bxid/.style={draw=idc,  fill=idc!12},
}

\begin{table}[t]
\centering
\normalsize
\setlength{\tabcolsep}{2pt}
\renewcommand{\arraystretch}{1.05}
\newcolumntype{L}[1]{>{\raggedright\arraybackslash\hsize=#1\hsize}X}
\begin{tabularx}{\columnwidth}{@{}L{0.86} L{1.09} L{1.05} c@{}}
\toprule
\textbf{Name} & \textbf{ID} & \textbf{OOD} & \textbf{Input Shape} \\
\midrule
Structured v.\ Random & IMDb~\cite{maas-EtAl:2011:ACL-HLT2011} & Random & 512 \\
NIH v.\ Enron & NIH Exporter~\cite{gao2020pile} & Enron Emails~\cite{gao2020pile} & 512 \\
Enron v.\ NIH & Enron Emails~\cite{gao2020pile} & NIH Exporter~\cite{gao2020pile} & 512 \\
NYT & 2018--2019 & 2020--2024 & 512 \\
ViT & ImageNet~\cite{deng2009imagenet} & Random & $224 {\times} 224 {\times} 3$ \\
\bottomrule
\end{tabularx}
\caption{\scheme{} evaluation datasets. ID and OOD are the in-distribution and
out-of-distribution corpora; the NYT rows give article publication years.
\emph{Input} is the sequence length in tokens, or the image resolution for
ViT. The NIH and Enron rows use a tokenizer trained on their own ID corpus;
all other rows use the stock tokenizer.}
\label{tab:datasets_table}
\end{table}

\paragraph{Statistical confirmation.}
\textbf{Figure~\ref{fig:token_distance_cdfs}} confirms the successive token-distance gap across all five
BERT model sizes: the in-distribution CDF rises sharply at small token distances, while the out-of-distribution CDF is consistently heavier-tailed.



\paragraph{Key insight.}
The processor is not reacting to input data semantics; it is reacting to the
\emph{geometry} of embedding memory accesses induced by the tokenizer.
All measured counters---fewer \ctr{page-table walk completions} for
members, fewer \ctr{off-core demand reads} for members, more
\ctr{reservation-station stalls} for members (because loads retire
faster and the back-end fills up)---are correlated projections of this
single root cause.
The difference is consistent across models, observable across hundreds
of counter events, and invisible in end-to-end timing.

\subsection{Algorithm}
\label{subsec:design}

\scheme{} is organized around four design principles:
\textbf{(P1)}~non-invasive measurement (the model is treated as a
fixed artifact);
\textbf{(P2)}~multi-witness evidence (263~counters organized across
seven subsystems form a jury, not a single gadget);
\textbf{(P3)}~controlled execution (core placement, SMT, frequency,
prefetcher state, and input normalization form a disclosed
configuration manifest);
\textbf{(P4)}~method separated from conclusion (output is
corroborating evidence, not a binary legal verdict).

The system proceeds in five stages:
\emph{(1)~artifact conditioning}---defining the provenance question
and constructing familiar/novel reference sets;
\emph{(2)~controlled execution harness}---applying hardware
configuration controls;
\emph{(3)~target-model inference}---executing on both reference sets
and the disputed artifact;
\emph{(4)~multi-witness footprint collection}---recording counter
telemetry, power, latency, and token locality;
\emph{(5)~analysis and evidence generation}---computing effect sizes,
subsystem grouping, and emitting an auditable report.

\subsection{Step-by-Step Attack Walkthrough: Example BERT}
\label{subsec:walkthrough_bert}

We provide an example of the \scheme{} attack on the 110.1\,M-parameter BERT-base
model trained for sentiment classification on 1{,}000 IMDB
reviews~\cite{maas-EtAl:2011:ACL-HLT2011}.

\paragraph{Step~1: Reference set construction.}
The familiar set $\mathcal{X}_F$ consists of 1{,}000 IMDB reviews from
the training split.
The novel set $\mathcal{X}_N$ consists of 1{,}000 sequences of
uniformly sampled token IDs from the BERT vocabulary, matched in
length.

\begin{figure}[!bp]
\centering
\begin{tikzpicture}
\begin{groupplot}[
  bxaxis,
  width=2.85cm, height=3.2cm,
  group style={group size=2 by 1, horizontal sep=1.35cm},
]

\nextgroupplot[
  ylabel={Cycles ($\times 10^{8}$)},
  ymin=4.975, ymax=5.262,
  ytick={5.00,5.05,5.10,5.15,5.20,5.25},
  yticklabel style={/pgf/number format/fixed, /pgf/number format/zerofill,
                    /pgf/number format/precision=2},
]
  \addplot[bxood, boxplot prepared={draw position=1,
    lower whisker=4.99093, lower quartile=5.08686, median=5.12761,
    upper quartile=5.15097, upper whisker=5.24708}] coordinates {};
  \addplot[bxid, boxplot prepared={draw position=2,
    lower whisker=4.99024, lower quartile=5.08614, median=5.12685,
    upper quartile=5.15021, upper whisker=5.24626}] coordinates {};

\nextgroupplot[
  ylabel={2M/4M page walks},
  ymin=386, ymax=556,
  ytick={400,440,480,520},
]
  \addplot[bxood, boxplot prepared={draw position=1,
    lower whisker=412, lower quartile=462, median=478,
    upper quartile=496, upper whisker=547}] coordinates {};
  \addplot[bxid, boxplot prepared={draw position=2,
    lower whisker=395, lower quartile=447, median=464,
    upper quartile=482, upper whisker=534}] coordinates {};

\end{groupplot}
\end{tikzpicture}
\caption{Per-inference cycle count (left) and 2M/4M-page table walks (right)
for in-distribution (ID) and out-of-distribution (OOD) inputs on 110M-parameter LLM, using the Structured v Random dataset in \textbf{Table~\ref{tab:datasets_table}}. Cycle counts are indistinguishable between the two groups
(medians differ by $7.6\times10^{4}$ cycles, $0.015\%$), but the page-walk
medians differ by 14 walks. 
}
\label{fig:introfig}
\end{figure}
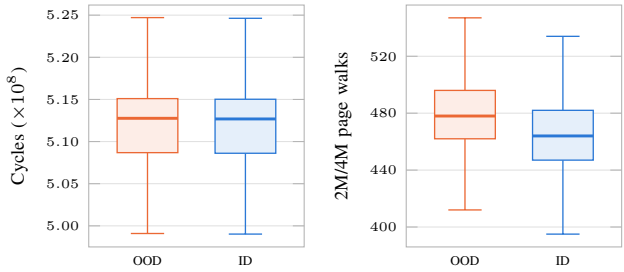
\paragraph{Step~2: Execution and counter collection.}
Each input is tokenized and padded to 512 tokens.
The model executes on a single core pinned via \texttt{taskset}, SMT
disabled, TurboBoost enabled, prefetchers at factory defaults. We collect all available performance counters on the system using PAPI~\cite{mucci1999papi}, plus RAPL energy. 

\paragraph{Step~3: Key observation.}
\textbf{Figure~\ref{fig:introfig} shows the result.}
End-to-end latency is statistically identical (${\sim}52$\,M cycles),
yet \ctr{page-table walk completions} are elevated by ${\sim}20$
walks for out-of-distribution inputs---consistent with the embedding-table locality
mechanism.

\paragraph{Step~4: MCF score computation.}
For each witness $w \in \mathcal{W}$, compute the $z$-score of the
disputed artifact relative to the familiar and novel reference
distributions.
Aggregate across witnesses to produce a subsystem-level summary and
overall MCF score.

\paragraph{Step~5: Evidence report.}
The final report includes per-counter effect sizes with 95\% bootstrap
CIs, a subsystem-level heatmap, the MCF score with estimated
false-positive rate, and the full configuration manifest.

\section{Evaluation}
\label{sec:eval}

\subsection{Experimental Methodology}
\label{sec:methodology}

\paragraph{Hardware configuration.}
All experiments run on a single isolated core pinned via
core affinity, with \texttt{isolcpus}
removing the core from the OS scheduler.
All experiments run with TurboBoost enabled (800\,MHz--4.1\,GHz).
SMT is disabled in all primary experiments. 
Counter access requires
\texttt{perf\_event\_paranoid\,=\,$-$1}
(\texttt{CAP\_PERFMON} on Linux~5.8+). Experiments marked IPEX use Intel Extensions for Pytorch, a Pytorch backend that makes use of Intel-specific features like AMX, rather than the default Pytorch CPU backend.

\paragraph{Models.}
We study three model architecture families spanning recent transformers:
\textbf{BERT} (4.4\,M, 11.3\,M, 29.1\,M,
41.7\,M, 110.1\,M parameters) fine-tuned for sentiment classification on the Large Movie Review Dataset~\cite{maas-EtAl:2011:ACL-HLT2011};
\textbf{ViT} (22\,M, 86\,M, 307\,M parameters) trained on ImageNet~\cite{deng2009imagenet} for image classification, and \textbf{Pythia} (70M, 160M, 410M, 1B, 1.4B, 2.8B parameters), a language model mimicking LLaMA~\cite{touvron2023llama} but with disclosed training data~\cite{biderman2023pythia}.

\paragraph{Datasets.}
Each model is paired with a corpus designed to isolate
training-exposure effects as seen in \textbf{Table~\ref{tab:datasets_table}}:
(i)~controlled-entropy images (ViT): CIFAR samples (ID)
interpolated with uniform entropy (OOD), decoupling input entropy from content semantics;
(ii)~semantically-structured vs.\ random tokens (BERT-Random): 1{,}000 IMDB reviews
(in-distribution) vs.\ 1{,}000 uniformly sampled random token sequences (out-of-distribution);
(iii)~Tokens from two different semantic contexts (BERT-NIH, BERT-Enron): 1{,}000 NIH Exporter samples (in-distribution) vs.\ 1{,}000 Enron Emails (out-of-distribution), and vice versa
(iv)~Pre/post-training-cutoff corpus (NYT): \emph{copyrighted} articles
before GPT's training cutoff (in-distribution) vs.\ post-cutoff
(out-of-distribution). All inputs are matched in length and size, ensuring that the same amount of work executes for each sample; therefore, any performance counter differences we find are due to how the model reacts to the input. 





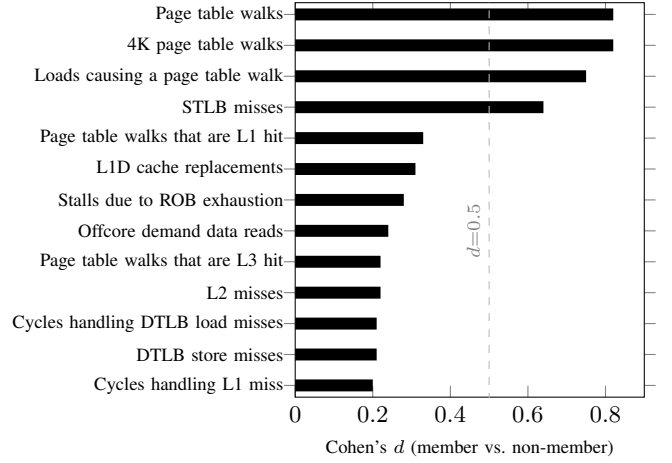
\begin{figure}[!htbp]
\centering
\begin{tikzpicture}
\begin{axis}[
  xbar,
  xmin=0, xmax=0.9,
  width=0.7\columnwidth,
  height=6.8cm,
  bar width=4.5pt,
  xlabel={Cohen's $d$ (member vs.\ non-member)},
  xlabel style={font=\scriptsize},
  xtick={0,0.2,0.4,0.6,0.8},
  xticklabel style={font=\small},
  ytick={0,...,12},
  yticklabels={
    {Cycles handling L1 miss},
    {DTLB store misses},
    {Cycles handling DTLB load misses},
    {L2 misses},
    {Page table walks that are L3 hit},
    {Offcore demand data reads},
    {Stalls due to ROB exhaustion},
    {L1D cache replacements},
    {Page table walks that are L1 hit},
    {STLB misses},
    {Loads causing a page table walk},
    {4K page table walks},
    {Page table walks}
  },
  yticklabel style={
    font=\scriptsize,
    text width=3.6cm,
    align=right,
    inner sep=0pt
  },
  enlarge y limits={abs=0.38},
  axis line style={thin},
  tick style={thin},
  clip=true,
]

\addplot[xbar,fill=black,draw=none] coordinates {
  (0.20,0)
  (0.21,1)
  (0.21,2)
  (0.22,3)
  (0.22,4)
  (0.24,5)
  (0.28,6)
  (0.31,7)
  (0.33,8)
  (0.64,9)
  (0.75,10)
  (0.82,11)
  (0.82,12)
};

\draw[dashed,gray!55,thin] (axis cs:0.5,-0.5) -- (axis cs:0.5,12.5);
\node[font=\scriptsize,gray,rotate=90,anchor=south]
  at (axis cs:0.5,5) {$d{=}0.5$};

\end{axis}
\end{tikzpicture}
\caption{Counters with Cohen's $d > 0.2$ for BERT-tiny (4.4M) on the Haswell microarchitecture, ranked by effect size.}
\label{fig:bert_filtered}
\end{figure}

\begin{figure}[!htbp]
\centering
    \begin{subfigure}{0.24\textwidth}
        \centering
        \includegraphics[width=\linewidth]{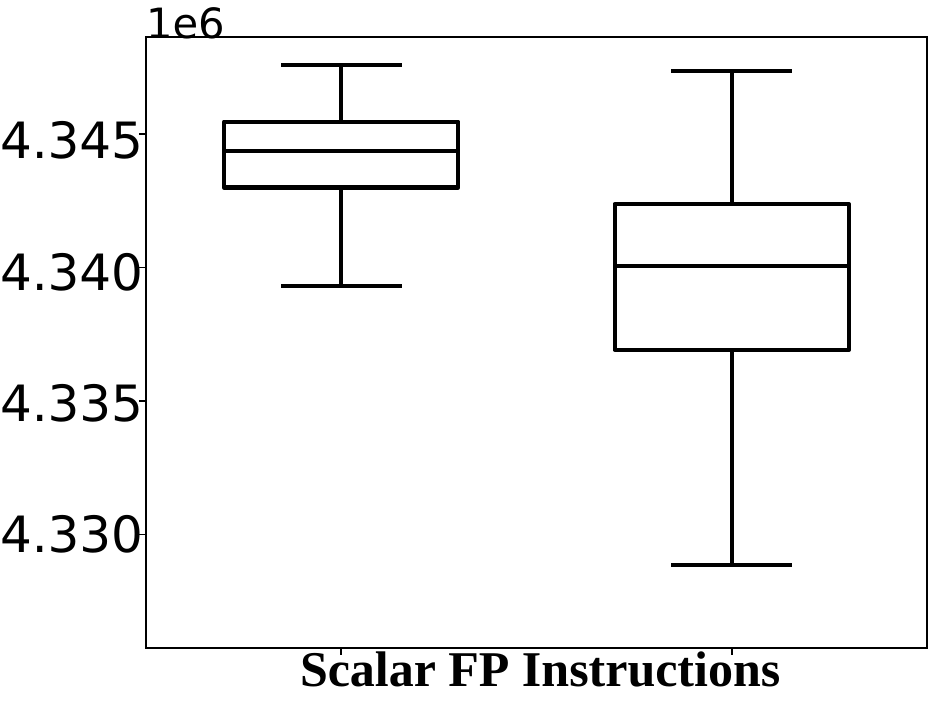}
    \end{subfigure}
    \hfill
    \begin{subfigure}{0.24\textwidth}
        \centering
        \includegraphics[width=\linewidth]{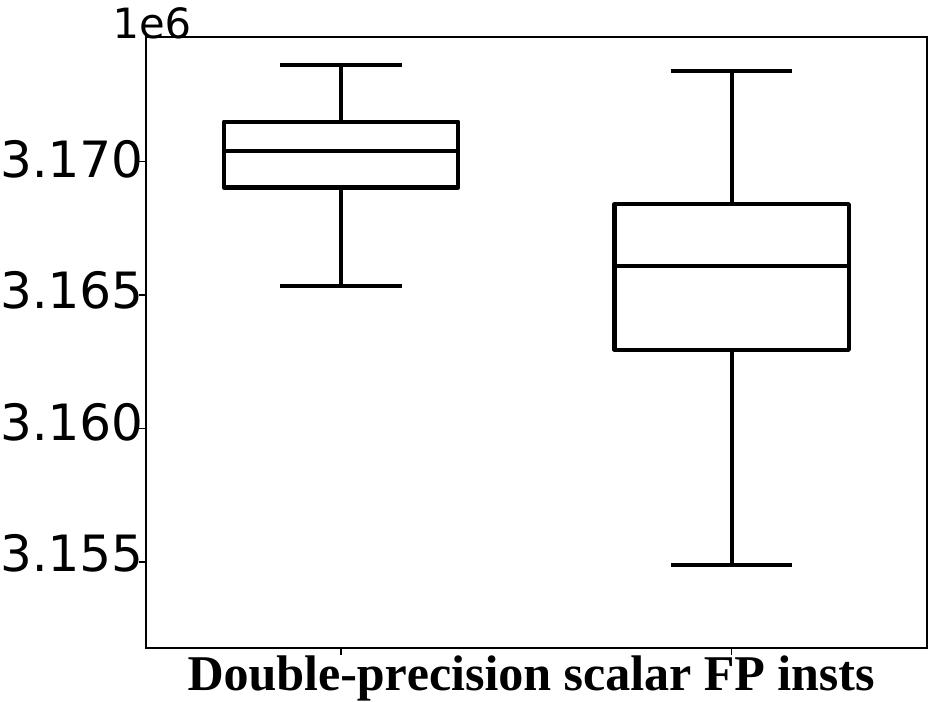}
    \end{subfigure}
    \hfill
    \begin{subfigure}{0.24\textwidth}
        \centering
        \includegraphics[width=\linewidth]{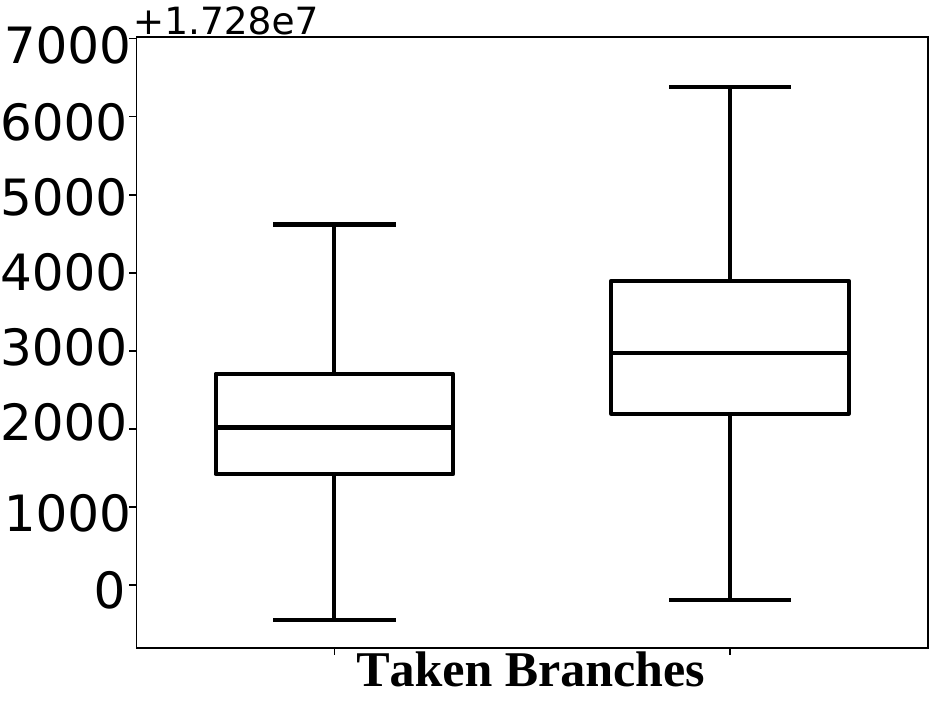}
    \end{subfigure}
    \hfill
    \begin{subfigure}{0.24\textwidth}
        \centering
        \includegraphics[width=\linewidth]{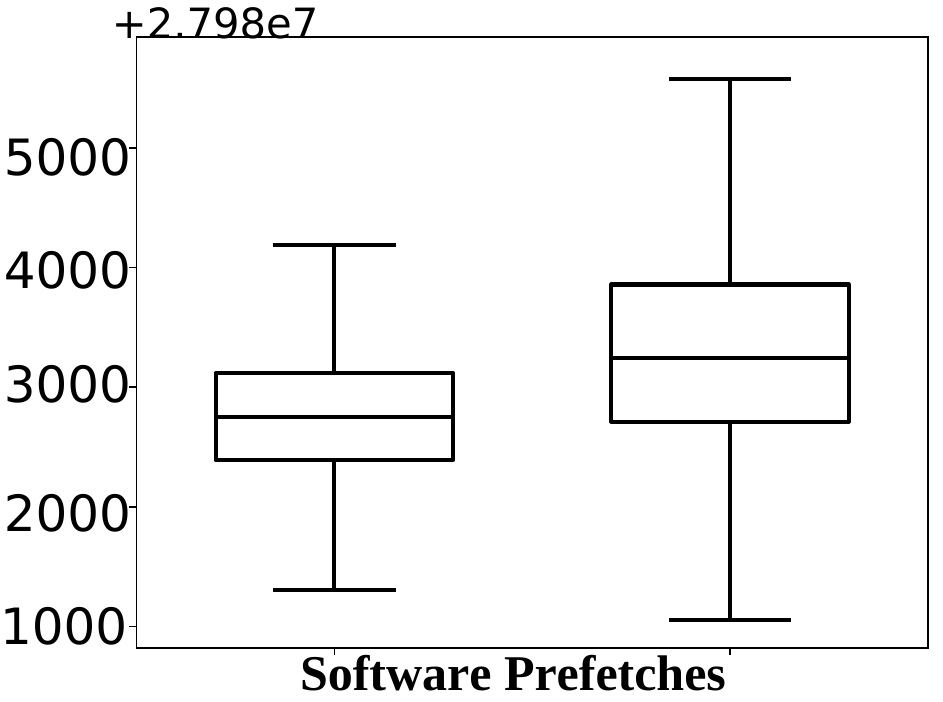}
    \end{subfigure}
    \caption{Four emergent counter signals for Vision Transformer on
    member (left) vs.\ nonmember (right) data.
    Despite identical clock-cycle counts, nonmembers retire more
    \ctr{scalar FP instructions} while members take more
    \ctr{conditional-taken branches} and trigger more
    \ctr{software prefetch accesses}---a previously unknown
    microarchitectural fingerprint in production vision transformers.}
    \label{fig:vit_emergent_counters}
\end{figure}

\subsection{Emergent Counter Rankings}
\label{sec:emergent}

We define a counter as \emph{emergent} if it exhibits medium-to-high
effect size ($d \geq 0.5$, $p \leq 0.15$) in the member-vs.-nonmember
comparison.
By convention, $d \geq 0.5$ is medium, $d \geq 0.8$ is large, and
$d \geq 1.5$ is exceptional.

\textbf{Figure~\ref{fig:bert_filtered}} ranks 13 emergent counters for
BERT-tiny (4.4\,M) on the Haswell microarchitecture. Counters relating to page-table walks exhibit emergent behavior and achieve the strongest separation between member and nonmember ($d=0.75$---$0.82$). The rest of the emergent counters are related to cache misses, like L1D cache replacement, ROB stalls, offcore data reads, L2 misses, and cycles handling L1 misses. \textbf{Figure~\ref{fig:vit_emergent_counters}} shows four emergent counters for Vision Transformers --- scalar FP instructions, double-precision FP instructions, taken branches, and software prefetches. Notably in the ViT and BERT cases, total cycle count and end-to-end timing are \emph{not} emergent, confirming that the leakage bypasses constant-time defenses. 

\begin{figure}[!htbp]
\centering
\begin{tikzpicture}
\begin{axis}[
  xbar,
  xmin=0, xmax=2.5,
  width=0.7\columnwidth,
  height=6.8cm,
  bar width=4.5pt,
  xlabel={Cohen's $d$ (member vs.\ non-member)},
  xlabel style={font=\scriptsize},
  xtick={0,0.25,0.5,0.75,1.0,1.25,1.50,1.75,2.0,2.25},
  xticklabel style={font=\scriptsize},
  ytick={1,...,15},
  yticklabels={
    {Port 0-dispatched FP instructions},
    {Microcode assists},
    {Port 1-dispatched FP instructions},
    {32-cycle stalled FE instructions},
    {Not-taken branches},
    {Instructions retired},
    {Instructions retired},
    {Memory instructions},
    {Loads},
    {Taken branches},
    {Conditional taken branches},
    {Software prefetches},
    {Double scalar FP instructions},
    {Scalar FP instructions},
    {FP microcode assists}
  },
  yticklabel style={
    font=\scriptsize,
    text width=3.6cm,
    align=right,
    inner sep=0pt
  },
  enlarge y limits={abs=0.38},
  axis line style={thin},
  tick style={thin},
  clip=true,
]

\addplot[xbar,fill=black,draw=none] coordinates {
  (0.25,1)
  (0.25,2)
  (0.29,3)
  (0.32,4)
  (0.35,5)
  (0.40,6)
  (0.42,7)
  (0.43,8)
  (0.59,9)
  (0.86,10)
  (0.99,11)
  (1.73,12)
  (1.72,13)
  (1.72,14)
  (2.52,15)
};

\draw[dashed,gray!55,thin] (axis cs:0.5,-0.5) -- (axis cs:0.5,12.5);
\node[font=\scriptsize,gray,rotate=90,anchor=south]
  at (axis cs:0.75,5) {$d{=}0.5$};

\end{axis}
\end{tikzpicture}
\caption{Top 15 effect size emergent counters for ViT-base (86M parameters) on the Sapphire Rapids microarchitecture, ranked by effect size. \ctr{FP microcode assists} ($d=2.52$) is the strongest single-counter membership signal reported to date.}
\label{fig:vit_emergent}
\end{figure}
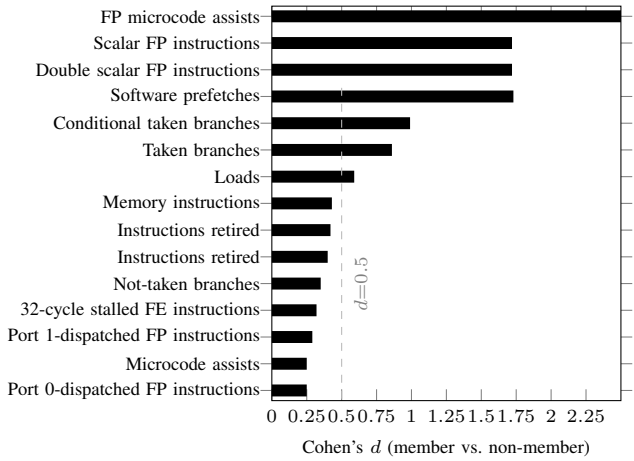

\textbf{Figure~\ref{fig:vit_emergent}} ranks the top 15 emergent counters for
ViT-base (86\,M) on Sapphire Rapids.
\ctr{Floating-point microcode
assists}\raw{ASSISTS.FP} dominates at $d = 2.52$---the largest
effect size across all model--architecture pairs in this study and the
largest hardware-based membership effect size reported to date.

\subsection{Power Side-Channel}
\label{subsec:power}

Per-inference socket power (Intel RAPL, SGX enabled) reveals a third
independent leakage channel orthogonal to both performance counters and
timing.
\textbf{Figure~\ref{fig:power_normalised}} shows power normalised to the
BERT-tiny member baseline.
The member/non-member gap grows monotonically from $+2.6\%$ at
BERT-mini to $+4.5\%$ at BERT-base, tracking model parameter count.
End-to-end wall-clock time is flat to within 1.2\% across all sizes,
confirming that timing defenses fail---yet RAPL power provides a
complementary and growing signal.



\subsection{BERT Member/Nonmember Classifier Study}
We collected the performance counter, energy, and power statistics for all BERT sizes on Sapphire Rapids across 100 samples of the IMDB dataset and 100 samples of verified non members of the same size which we generated with 10 repeats per sample and stacked the resulting statistics into a dataset of 2000 samples each with 1,596 features. We trained a 4-layer multilayer perceptron classifier on the the resultant data using an 80/20 train/test split, with the layers having 512, 128, 32, and one 
neuron, respectively. We trained for 10 epochs with a dropout of 0.4 in the first 2 layers. \textbf{Table~\ref{tab:performance}} contains our results. In most cases, we achieved an accuracy of over 90\% in determining IMDB samples vs random sequences using only performance counters, regardless of model size.

Similarly, we repeated the Structured v Random experiment with BERT running on Intel Extensions for Pytorch (IPEX), a backend using Intel AMX and Intel AVX-512-VNNI kernels representing modern accelerator hardware. Interestingly, we see a clear monotonic increase in AUC and accuracy for increasing model size when using the IPEX backend, implying the use of modern accelerator hardware makes \scheme{} more effective at larger model sizes. 

Finally, we repeat the methodology except using the Enron v NIH and NIH v Enron datasets to see if \scheme{} can distinguish between two nonuniform distributions. We find that despite both in-distribution and out-of-distribution texts being valid sentences, \scheme{} is still easily able to separate member and non-member with 98\% accuracy. 

\begin{figure}[!b]
\centering
\includegraphics[width=\columnwidth]{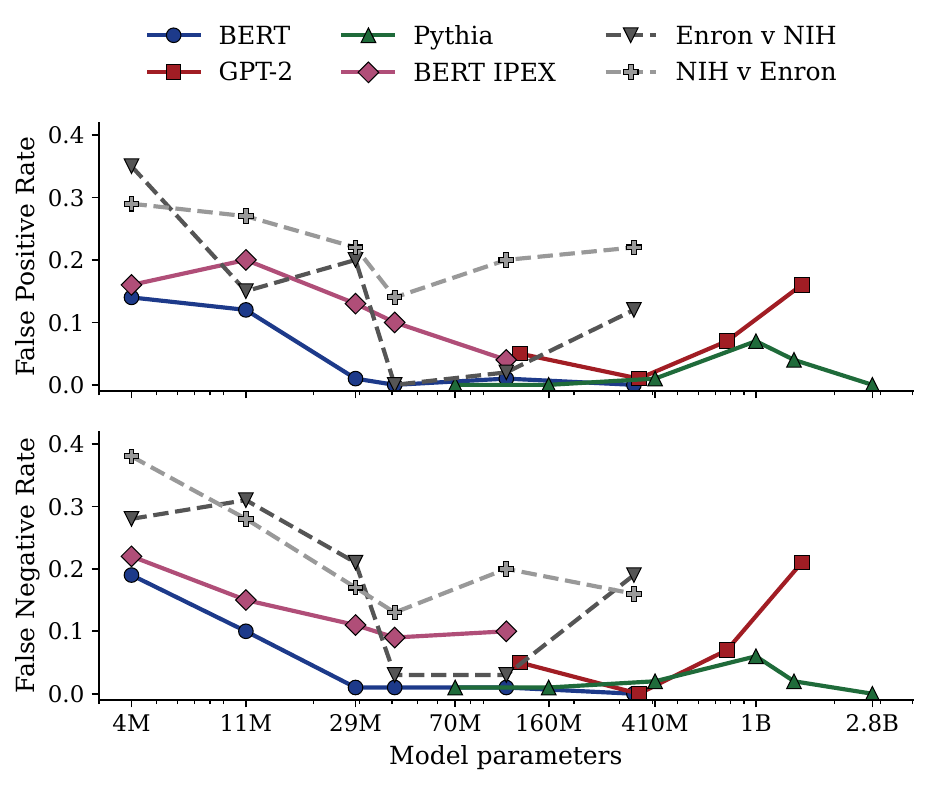}
\vspace{-6pt}
\caption{\textbf{False positive and false negative rates vs.\ model size.}
Both error rates drop sharply as model size increases for BERT and
Pythia , while GPT-2 reverses at 1.5\,B
parameters. Cross-domain pairs remain elevated across
all sizes.}
\label{fig:fpr-fnr}
\end{figure}

As \textbf{Figure~\ref{fig:fpr-fnr}} shows, both the false-positive and
false-negative rates drop to near zero once the model exceeds
${\sim}30$\,M parameters for BERT and ${\sim}70$\,M for Pythia,
confirming that the attack signal strengthens with capacity.

\subsection{New York Times Articles Study}
\begin{figure}[!htbp]
\centering

\definecolor{y2019}{HTML}{D83B01}
\definecolor{y2020}{HTML}{0078D4}
\definecolor{y2021}{HTML}{107C10}
\definecolor{y2022}{HTML}{5C2D91}
\definecolor{y2023}{HTML}{008272}

\begin{tikzpicture}[baseline]
  \node[draw=gray!60, line width=.4pt, rounded corners=2pt, inner sep=3pt] {%
    \begin{tikzpicture}[baseline=-0.6ex, every node/.style={inner sep=1pt}]
      \node at (0,0) {\tikz{\fill[y2019] (0,0) circle (2.2pt);}};
      \node[anchor=west] at (0.13,0) {2019};
      \node at (1.05,0) {\tikz{\draw[y2020,line width=.9pt] (-2pt,-2pt) rectangle (2pt,2pt);}};
      \node[anchor=west] at (1.18,0) {2020};
      \node at (2.10,0) {\tikz{\draw[y2021,line width=.9pt] (-2.3pt,-1.8pt)--(2.3pt,-1.8pt)--(0,2.4pt)--cycle;}};
      \node[anchor=west] at (2.23,0) {2021};
      \node at (3.15,0) {\tikz{\draw[y2022,line width=.9pt] (0,2.6pt)--(2.1pt,0)--(0,-2.6pt)--(-2.1pt,0)--cycle;}};
      \node[anchor=west] at (3.28,0) {2022};
      \node at (4.20,0) {\tikz{\draw[y2023,line width=.9pt] (-2.3pt,1.8pt)--(2.3pt,1.8pt)--(0,-2.4pt)--cycle;}};
      \node[anchor=west] at (4.33,0) {2023};
    \end{tikzpicture}};
\end{tikzpicture}

\vspace{2pt}

\begin{tikzpicture}
\pgfplotsset{
  nytpanel/.style={
    width=6.7cm, height=2.3cm, scale only axis,
    xmin=2018.55, xmax=2023.45,
    xtick={2019,2020,2021,2022,2023},
    xticklabels={2019,2020,2021,2022,2023},   
    label style={font=\normalsize}, tick label style={font=\normalsize},
    title style={font=\normalsize, yshift=-2pt},
    axis line style={draw=gray!60}, tick style={draw=gray!60},
    ymajorgrids, grid style={line width=.2pt, draw=gray!25},
    scaled y ticks=false,
    clip=false,
  },
  p2019/.style={only marks, mark=*,          mark size=2.1pt, color=y2019},
  p2020/.style={only marks, mark=square,     mark size=2.1pt, line width=.9pt, color=y2020},
  p2021/.style={only marks, mark=triangle,   mark size=2.6pt, line width=.9pt, color=y2021},
  p2022/.style={only marks, mark=diamond,    mark size=2.6pt, line width=.9pt, color=y2022},
  p2023/.style={only marks, mark=triangle,   mark size=2.6pt, line width=.9pt, color=y2023,
                mark options={rotate=180}},
  meanline/.style={color=black!80, line width=.8pt, mark=none},
  refline/.style={dashed, color=gray!70, line width=.6pt, forget plot},
}
\newcommand{\band}[2]{%
  \fill[gray!15] (axis cs:2018.55,#1) rectangle (axis cs:2023.45,#2);}

\begin{groupplot}[
  nytpanel,
  group style={group size=1 by 3, vertical sep=0.72cm,
               x descriptions at=edge bottom},
]

\nextgroupplot[
  title={(a) \texttt{OFFCORE\_REQUESTS.DEMAND\_DATA\_RD}},
  ylabel={Reads ($\times10^{-3}$)},
  ymin=2.30, ymax=7.55, ytick={3,4.5,6,7.5},
  yticklabel style={/pgf/number format/fixed,
                    /pgf/number format/zerofill,
                    /pgf/number format/precision=1},
]
  \band{6.67}{7.32}
  \addplot[refline] coordinates {(2018.55,6.98) (2023.45,6.98)};
  \addplot[meanline] coordinates {(2019,3.048) (2020,6.970) (2021,6.960)
                                  (2022,6.970) (2023,7.015)};
  \addplot[p2019] coordinates {(2019,2.53) (2019,3.05) (2019,3.22) (2019,3.39)};
  \addplot[p2020] coordinates {(2020,6.87) (2020,7.07)};
  \addplot[p2021] coordinates {(2021,6.91) (2021,7.01)};
  \addplot[p2022] coordinates {(2022,6.88) (2022,7.06)};
  \addplot[p2023] coordinates {(2023,6.88) (2023,7.15)};
  \draw[<->, line width=.6pt] (axis cs:2019.42,3.05) -- (axis cs:2019.42,6.98);
  \node[anchor=west, font=\normalsize] at (axis cs:2019.52,4.55) {$-56\%$};

\nextgroupplot[
  title={(b) \texttt{DTLB\_LOAD\_MISSES.WALK\_PENDING}},
  ylabel={Walk cycles (\%)},
  ymin=5.30, ymax=7.05, ytick={5.6,6.0,6.4,6.8},
  yticklabel style={/pgf/number format/fixed,
                    /pgf/number format/zerofill,
                    /pgf/number format/precision=1},
]
  \band{6.31}{7.03}
  \addplot[refline] coordinates {(2018.55,6.65) (2023.45,6.65)};
  \addplot[meanline] coordinates {(2019,5.840) (2020,6.615) (2021,6.650)
                                  (2022,6.760) (2023,6.583)};
  \addplot[p2019] coordinates {(2019,5.40) (2019,5.72) (2019,5.84) (2019,6.40)};
  \addplot[p2020] coordinates {(2020,6.58) (2020,6.65)};
  \addplot[p2021] coordinates {(2021,6.48) (2021,6.60) (2021,6.87)};
  \addplot[p2022] coordinates {(2022,6.65) (2022,6.75) (2022,6.88)};
  \addplot[p2023] coordinates {(2023,6.44) (2023,6.55) (2023,6.76)};
  \draw[<->, line width=.6pt] (axis cs:2019.42,5.84) -- (axis cs:2019.42,6.65);
  \node[anchor=west, font=\normalsize] at (axis cs:2019.52,6.04) {$-12\%$};

\nextgroupplot[
  title={(c) End-to-end timing},
  ylabel={Cycles ($\times10^{10}$)},
  xlabel={NYT article publication year},
  ymin=1.015, ymax=1.400, ytick={1.04,1.12,1.20,1.28,1.36},
  yticklabel style={/pgf/number format/fixed,
                    /pgf/number format/zerofill,
                    /pgf/number format/precision=2},
]
  \band{1.140}{1.181}
  \addplot[refline] coordinates {(2018.55,1.181) (2023.45,1.181)};
  \addplot[meanline] coordinates {(2019,1.166) (2020,1.185) (2021,1.173)
                                  (2022,1.184) (2023,1.182)};
  \addplot[p2019] coordinates {(2019,1.036) (2019,1.088) (2019,1.166) (2019,1.375)};
  \addplot[p2020] coordinates {(2020,1.098) (2020,1.112) (2020,1.345)};
  \addplot[p2021] coordinates {(2021,1.095) (2021,1.119) (2021,1.305)};
  \addplot[p2022] coordinates {(2022,1.097) (2022,1.110) (2022,1.346)};
  \addplot[p2023] coordinates {(2023,1.076) (2023,1.112) (2023,1.358)};
  \node[anchor=north, font=\normalsize, align=center, fill=white,
        draw=gray!60, line width=.4pt, rounded corners=2pt, inner sep=2pt]
    at (axis cs:2021.05,1.268) {Constant time: max diff.\ 1.2\%};

\end{groupplot}
\end{tikzpicture}


    \caption{NYT article inference on unmodified GPT\,/\,Sapphire Rapids,
    partitioned by publication year (2019\,$=$\, potential member;
    2020--2023\,$=$\,non-member).
    \textbf{(a)}~\ctr{Off-core demand reads} drop sharply for 2019
    (${\sim}90\%$ lower).
    \textbf{(b)}~\ctr{TLB walk-pending cycles} show a parallel
    $12\%$ reduction for 2019.
    \textbf{(c)}~End-to-end timing remains constant to within
    $1.2\%$---constant-time defenses are blind to this channel.}
    \label{fig:nyt_overview}
\label{fig:nyt_years}
\end{figure}
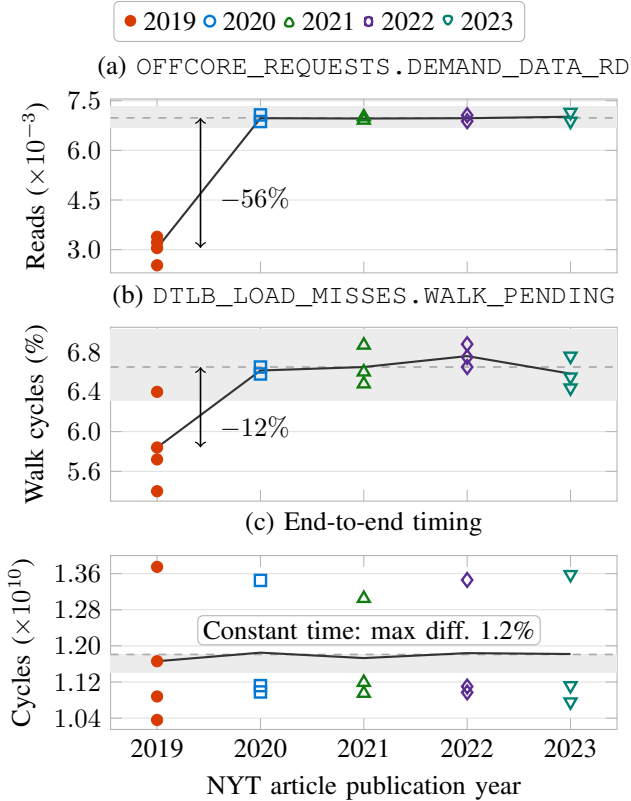

We collected the performance counter and timing statistics of running GPT on New York Times articles from 2019, 2020, 2021, 2022, and 2023. GPT's training cutoff was November 2019; hence, the article from 2019 is a possible member, or at least in-distribution, to GPT's training set whereas the articles from 2020-2023 can not be. \textbf{Figure~\ref{fig:nyt_overview}} shows the result of this study, finding that the 2019 article exhibits 56\% fewer off-core demand reads over the 2020-2023 articles and 12\% fewer DTLB load misses, all while remaining within 1\% of runtime bypassing constant-time defenses. Therefore, GPT is more familiar with the 2019 article than the other articles, fitting what we know about the training date cutoff.


\begin{table*}[t]
\centering
\setlength{\tabcolsep}{3.5pt}
\renewcommand{\arraystretch}{1.05}
\footnotesize
\begin{tabular}{@{}lrrrr@{\hspace{13pt}}lrrrr@{\hspace{13pt}}lrrrr@{}}
\toprule
\multicolumn{5}{@{}l}{\textbf{BERT~\cite{devlin2018bert}} — Structured v Random} &
\multicolumn{5}{@{}l}{\textbf{GPT-2~\cite{radford2019language}} — Structured v Random} &
\multicolumn{5}{@{}l}{\textbf{Pythia~\cite{biderman2023pythia}} — Structured v Random} \\
\cmidrule(r{11pt}){1-5}\cmidrule(r{11pt}){6-10}\cmidrule{11-15}
Params & AUC & Acc. & FPR & FNR &
Params & AUC & Acc. & FPR & FNR &
Params & AUC & Acc. & FPR & FNR \\
\midrule
4M   & 0.92 & 84\%  & 0.14 & 0.19 & 124M & 0.99 & 96\%  & 0.05 & 0.05 & 70M   & 1.00 & 99\%  & 0.00 & 0.01 \\
11M  & 0.95 & 89\%  & 0.12 & 0.10 & 355M & 1.00 & 100\% & 0.01 & 0.00 & 160M  & 1.00 & 99\%  & 0.00 & 0.01 \\
29M  & 0.99 & 100\% & 0.01 & 0.01 & 774M & 0.98 & 93\%  & 0.07 & 0.07 & 410M  & 0.99 & 98\%  & 0.01 & 0.02 \\
41M  & 0.99 & 100\% & 0.00 & 0.01 & 1.5B & 0.90 & 82\%  & 0.16 & 0.21 & 1B    & 0.98 & 94\%  & 0.07 & 0.06 \\
110M & 0.99 & 99\%  & 0.01 & 0.01 & --   & --   & --    & --   & --   & 1.4B  & 0.99 & 97\%  & 0.04 & 0.02 \\
340M & 1.00 & 100\% & 0.00 & 0.00 & --   & --   & --    & --   & --   & 2.8B  & 1.00 & 99\%  & 0.00 & 0.00 \\
\midrule
\multicolumn{5}{@{}l}{\textbf{BERT IPEX~\cite{devlin2018bert}} — Structured v Random} &
\multicolumn{5}{@{}l}{\textbf{BERT~\cite{devlin2018bert}} — Enron v NIH} &
\multicolumn{5}{@{}l}{\textbf{BERT~\cite{devlin2018bert}} — NIH v Enron} \\
\cmidrule(r{11pt}){1-5}\cmidrule(r{11pt}){6-10}\cmidrule{11-15}
Params & AUC & Acc. & FPR & FNR &
Params & AUC & Acc. & FPR & FNR &
Params & AUC & Acc. & FPR & FNR \\
\midrule
4M   & 0.89 & 81\% & 0.16 & 0.22 & 4M   & 0.75 & 69\% & 0.35 & 0.28 & 4M   & 0.73 & 67\% & 0.29 & 0.38 \\
11M  & 0.91 & 83\% & 0.20 & 0.15 & 11M  & 0.84 & 77\% & 0.15 & 0.31 & 11M  & 0.80 & 73\% & 0.27 & 0.28 \\
29M  & 0.95 & 88\% & 0.13 & 0.11 & 29M  & 0.88 & 80\% & 0.20 & 0.21 & 29M  & 0.89 & 80\% & 0.22 & 0.17 \\
41M  & 0.96 & 90\% & 0.10 & 0.09 & 41M  & 1.00 & 98\% & 0.00 & 0.03 & 41M  & 0.87 & 94\% & 0.14 & 0.13 \\
110M & 0.98 & 93\% & 0.04 & 0.10 & 110M & 1.00 & 97\% & 0.02 & 0.03 & 110M & 0.88 & 80\% & 0.20 & 0.20 \\
--   & --   & --   & --   & --   & 340M & 0.92 & 84\% & 0.12 & 0.19 & 340M & 0.90 & 81\% & 0.22 & 0.16 \\
\bottomrule
\end{tabular}
\caption{\scheme{} attack statistics on production transformers.}
\label{tab:performance}
\end{table*}

\section{Microarchitectural Insights and Redesign Guidance}
\label{sec:arch}
This paper does more than introduce a new family of side channels.
Its deeper contribution is \emph{architectural}: it provides one of
the first detailed CPU-side characterisations of how production CNNs,
LLMs, and vision transformers stress the microarchitecture, how that
stress changes with input familiarity, and how it evolves as model
size scales.
Across more than 290 hardware performance counters, along with
latency, energy, and power, we show that training influence
does not remain confined to model outputs---it reshapes the execution
footprint of the processor itself.
Production transformers do not stress the processor uniformly:
training predictability and model scale reshape branch behavior,
front-end delivery, TLB activity, cache locality, reservation-station
occupancy, execution-port pressure, power, and latency across the
full microarchitecture pipeline.

These results therefore matter not only to privacy and security
researchers, but also to \emph{performance architects and hardware
designers}, because each counter disparity is simultaneously a
leakage signal and a clue about where modern AI workloads stress,
underutilise, or destabilise the machine.


\paragraph{Translation structures.}
\ctr{Page-table walk completions for 2\,MB/4\,MB pages} increase with
BERT model size and separate familiar from novel inputs across all
five parameter points.
Transformer embedding-table access patterns stress the page-table
walker far more than prior workloads---300+ walk completions per
inference, each costing hundreds of
cycles~\cite{loughlin2021dolma} 
This motivates embedding-aware TLB partitioning, large-page-backed
embedding tables, and speculative page-walk prefetch tuned to
gather/scatter access patterns.

\paragraph{Speculation recovery.}
\ctr{Pipeline clears} ($d \approx 0.84$) and \ctr{pipeline recovery
cycles} ($d \approx 0.73$) differ between familiar and novel inputs
even when wall-clock time does not---further evidence that the leakage
operates below the timing floor and bypasses constant-time defenses.
Future CPUs targeting LLM inference could benefit from ``ML-phase''
hints that adapt branch-predictor confidence thresholds for
out-of-distribution inputs.

\paragraph{Cache hierarchy and prefetch.}
Embedding-table scatter drives L3 miss counts and \ctr{off-core demand
reads} for novel inputs; familiar inputs enjoy greater cache residency
because the model has been conditioned to their token distribution.
Prefetcher designs that recognize embedding-table stride
signatures could reduce both miss rates and the locality-driven
leakage that \scheme{} exploits.

\paragraph{Scheduler pressure.}
\ctr{Reservation-station empty cycles} and \ctr{store-buffer stalls}
vary with both model size and input familiarity, revealing phases
where speculative work and memory-latency hiding interact with
novelty-driven traffic.

\paragraph{Power and energy.}
Per-inference power differs between familiar and novel inputs even when
latency does not, constituting a leakage channel orthogonal to
timing~\cite{Platypus2021} as shown in \textbf{Figure~\ref{fig:power_normalised}}.
Increasing training-set coverage reduces average inference energy;
``training coverage'' becomes a new axis in energy modeling of
production LLM deployments.

\section{Related Work}
\label{sec:related}

\paragraph{Software-based membership inference.}
The foundational MIA by Shokri et al.~\cite{Shokri2017SP} used shadow
models to infer membership from classification outputs.
LiRA~\cite{carlini2022membership} improved attack power through
likelihood-ratio tests.
Label-only attacks~\cite{choo2021membership} showed that
differential privacy~\cite{Abadi2016DP} is the only effective
defense.
Subsequent work extended MIAs to recommender
systems~\cite{zhang2021membership}, federated
learning~\cite{chang2024efficient}, in-context
learning~\cite{wen2024membership}, vision-language
models~\cite{hu2025membership}, and
LLMs~\cite{rossi2025membership}.
All prior works require model outputs, confidence scores, running an expensive surrogate model, or multiple inferences; \scheme{} is the first cheap, one-shot out-of-distribution detector that does not require any model output or surrogate model.

\paragraph{Hardware-assisted privacy attacks.}
GateBleed~\cite{kalyanapu2025gatebleed} exploits Intel AMX
power-gating timing for MIA on adaptive networks.
FeatureBleed~\cite{asher2026featurebleed} infers private attributes
from sparsity-optimized accelerators.
Timing-channel attacks~\cite{Akinsanya2024TimingCI} target adaptive
neural networks with observable latency variation.
\scheme{} is the first to demonstrate membership leakage on
\emph{static, branchless, production} networks where timing is
constant and all prior hardware-assisted attacks fail.

\paragraph{Property inference.}
Property inference attacks~\cite{ganju2018property} determine
undisclosed properties of training data.
Extensions cover GANs~\cite{zhou2021property},
CNNs~\cite{parisot2021property}, graph neural
networks~\cite{wang2022group}, and federated
learning~\cite{wang2022poisoning}.
\scheme{} subsumes property inference as a special case: when the
``property'' is membership in a temporal slice of a publisher's
archive, our hardware footprint provides direct evidence.

\paragraph{Microarchitectural side channels.}
Prior work establishes that timing channels persist on formally
verified kernels~\cite{cock2014last} and surveys the attack
surface exhaustively~\cite{ge2018survey}.
Our finding that 15+ counter events discriminate training members from
nonmembers \emph{while timing remains flat} adds a new entry to that
taxonomy: a \emph{software-optimization-mediated counter channel} that
operates below the timing floor.
Our 290+ event trace dataset enables the community to compute formal
channel-capacity bounds for each leaking event across three model
families and two microarchitectures.

\paragraph{Microarchitectural characterization of AI workloads.}
While this paper introduces a new series of side-channel techniques,
its broader contribution is to expose the \emph{microarchitectural
anatomy} of modern AI execution on CPUs.
Prior CPU characterization studies have focused on traditional server
workloads; no prior work has systematically profiled how training
familiarity, input entropy, and model scale reshape the full
microarchitecture pipeline---from branch prediction through TLB
activity and cache locality to execution-port utilization and power.
This paper is therefore not only a security result; it is also a
microarchitectural characterization study of production AI workloads,
with direct implications for future processor optimization and
hardware design.
Our traces show that transformer embedding lookups generate 300+
page-table walk completions per inference on Sapphire Rapids, each
costing hundreds of cycles---a bottleneck invisible to prior workload
characterization and one that motivates embedding-aware TLB
partitioning and large-page-backed embedding table layouts.

\section{Conclusion}
\label{sec:conclusion}
Membership inference is the root of privacy in modern AI. This paper establishes that hardware detectors can bypass state-of-the-art software defenses in constant time models, and alarmingly, these hardware-level leakage signals amplify as models—such as LLMs, Vision Transformers, and CNNs—scale and as new accelerators  gets integrated in CPUS. Consequently, user privacy is fundamentally vulnerable at the hardware level. As AI becomes ubiquitous across all computing domains, and hardware accelerators becoming more integrated, relying solely on software defenses is no longer sufficient. \scheme{} takes the first step in analyzing how these privacy-compromising hardware signals manifest as we relentlessly scale and deploy these models or integrate aggressive on core acceleration for AI.

\bibliographystyle{IEEEtran}
\bibliography{refs}

\end{document}